\documentclass[11pt]{article}

\usepackage{jheppub} 
\usepackage{calc}
\usepackage[running]{lineno}
\RequirePackage{graphicx}
\usepackage{placeins}    
\usepackage{amsmath,amssymb}
\usepackage{mathtools}
\usepackage{array}
\usepackage{amssymb}
\usepackage{float}
\usepackage{subcaption}
\usepackage[percent]{overpic}
\usepackage{xfrac}
\usepackage{physics}

\usepackage{xcolor}
\usepackage{bigints}
\usepackage{comment}
\usepackage[T1]{fontenc}

\newcommand{\sd}{\text{d}}
\newcommand{\zc}{z_{\text{cut}}}
\newcommand{\tilp}{\{\tilde{p}\}}

\usepackage{color}
\definecolor{darkgreen}{rgb}{0,0.5,0}
\definecolor{darkblue}{rgb}{0,0,0.7}
\definecolor{darkred}{rgb}{0.5,0,0.0}
\definecolor{darkorange}{rgb}{0.8,0.4,0.0}

\makeatletter
\g@addto@macro\bfseries{\boldmath}
\makeatother
\title{QCD resummation for groomed jet observables at NNLL+NLO}
\preprint{CERN-TH-2022-170}
\author[a,b]{Mrinal Dasgupta,}
\author[a]{Basem Kamal El-Menoufi}
\author[c]{and Jack Helliwell}
\affiliation[a]{Lancaster-Manchester-Sheffield Consortium for Fundamental Physics, School of Physics
  \& Astronomy, University of Manchester, Manchester M13 9PL, United Kingdom}
  \affiliation[b]{CERN, Theoretical Physics Department, CH-1211, Geneva 23, Switzerland}
\affiliation[c]{Rudolf Peierls Centre for Theoretical Physics, Parks Road, Oxford OX1 3PU, UK}
\emailAdd{mrinal.dasgupta@manchester.ac.uk}
\emailAdd{basem.el-menoufi@manchester.ac.uk}
\emailAdd{jack.helliwell@physics.ox.ac.uk}


\abstract{We use a direct QCD approach to carry out the next-to-next-to-leading logarithmic (NNLL) resummation for observables groomed with the modified mass-drop tagger (Soft Drop $\beta=0$).  We focus on observables which are additive given an arbitrary number of soft-collinear emissions. For this class of observables, we arrange the structure of the NNLL terms into two distinct categories. The first defines a simplified {\em inclusive} tagger, whereby the NNLL collinear structure is directly related to ungroomed observables.  The second defines a clustering correction which takes a particularly simple form when the Cambridge-Aachen (C/A) algorithm is used to cluster the jets. We provide, in addition to the QCD resummation of groomed jet mass, the first NNLL resummed predictions, matched to NLO, for a range of groomed jet angularities with mMDT grooming. Moreover, we also include for the first time in the same calculation, finite $\zc$ effects computed at NLL level alongside the small $\zc$ NNLL results which simultaneously improves upon both of the calculations used for groomed jet mass phenomenological studies to date. While for simplicity we focus on $e^{+}e^{-}$ collisions, the essential NNLL resummation we develop is process independent and hence with the appropriate NLO matching our results are also applicable for hadron collider phenomenology.
}

\begin{document}

\maketitle

\section{Introduction}\label{sec:mMDTResumIntro}

Grooming has emerged as an essential technique for jet substructure studies at the LHC where the high levels of pileup (PU) and underlying event (UE) lead to large amounts of soft QCD radiation originating from outside the jet being clustered into jets, masking their intrinsic substructure. Grooming was initially conceived as a method for removing some of this soft radiation from jets so as to enhance the resolution with which the substructure could be examined. One of the most popular grooming algorithms is mMDT \cite{Dasgupta:2013ihk} or equivalently Soft Drop \cite{Larkoski:2014wba} with $\beta=0$. Aside from reducing the effects of UE and PU, it was also found that grooming jets with mMDT reduces the size of hadronisation corrections to observables measured on those jets \cite{Dasgupta:2013ihk}. This naturally makes observables computed on groomed jets good candidates for direct comparison between perturbative QCD calculations and experimental measurements, or extractions of $\alpha_s$ from jet substructure measurements through fitting precision calculations to data \cite{Proceedings:2018jsb,Marzani:2019evv,Hannesdottir:2022rsl}. On top of this, it was found that grooming with mMDT removes any non-global logarithms (NGLs)\cite{Dasgupta:2001sh,Dasgupta:2013ihk} which would be present for the same observable computed on un-groomed jets. This elimination of NGLs removes one of the major difficulties associated with producing precise resummed predictions for jet shape observables. Because of these favourable properties a number of calculations have been carried out for groomed observables and compared directly to unfolded LHC data \cite{Caletti:2021oor,Reichelt:2021svh,CMS:2018vzn,CMS:2021vsp,Marzani:2017mva,Frye:2016aiz,ATLAS:2020bbn,ATLAS:2019mgf}.

There are a number of approaches to producing all orders predictions for groomed jet observables including Monte Carlo event generators, which typically offer limited logarithmic accuracy \cite{Dasgupta:2018nvj}, and analytic resummations carried out using either perturbative QCD methods or soft collinear effective theory (SCET). The resummation for the groomed jet mass was first carried out in \cite{Dasgupta:2013ihk} at NLL accuracy.\footnote{In this work we adopt a logarithmic counting scheme where the leading logarithms are double logs despite these being absent for observables groomed with mMDT.} This resummation was matched to NLO \cite{Marzani:2017mva} and compared with unfolded LHC data in \cite{CMS:2018vzn} which also showed a comparison with an NNLL calculation carried out in the small $\zc$ limit \cite{Frye:2016aiz,Bell:2018oqa,Bell:2018jvf} without NLO matching. This latter calculation was carried out in the SCET formalism and has since been extended to N$^3$LL accuracy by extracting the relevant anomalous dimensions from fixed-order codes \cite{Kardos:2020ppl}, though the N$^3$LL resummation has not been compared to data. Although the SCET factorisation theorem has been presented for multiple observables, the required anomalous dimensions are not known to NNLL accuracy other than for the jet mass or equivalent observables, as far as we are aware \cite{Frye:2016aiz,Kang:2018vgn}. Another approach to producing resummed predictions, in the absence of grooming, is to use the \texttt{CAESAR} and \texttt{ARES} programs to generate, respectively,  NLL and NNLL predictions \cite{Banfi:2004yd,Banfi:2014sua}. Never the less, the \texttt{CAESAR} plug-in for Sherpa \cite{Gerwick:2014gya} was applied in Refs.~\cite{Caletti:2021oor,Reichelt:2021svh} to produce NLL predictions for groomed jet angularities, where the need for NNLL resummation to reduce the uncertainty was noted. Beyond NLL accuracy, the formal lack of recursive infra-red and collinear safety (rIRC safety) of groomed observables starts to have an impact, thus precluding direct use of the ARES formalism to generate NNLL predictions for groomed observables and motivating the analytic resummation carried out in this work.

In this work we extend the NLL resummation carried out in \cite{Dasgupta:2013ihk} to reach NNLL accuracy. Previously a fixed-order study of the mMDT heavy-jet mass was carried out in \cite{Anderle:2020mxj} (see also \cite{Dasgupta:2021hbh}), which enabled us to uncover the relation between the NNLL terms, in the presence of grooming, with the equivalent structures known from resummed results for generic ungroomed two-jet observables \cite{Banfi:2014sua,Banfi:2018mcq}. In this work we build on these insights to obtain the resummed distribution for a wide class of rIRC safe observables, which are additive in the presence of an arbitrary number of soft and collinear emissions, whose momenta are denoted by $k_i$,
\begin{equation}
	V_{\mathrm{s.c}}(\tilp,k_1,...,k_n)=\sum_{i=1}^n V_{\mathrm{s.c}}(\tilp,k_i)\ ,
\end{equation}
where $\tilp$ denotes the set of primary hard partons in the final state. We focus in this paper on $e^+ e^-$ collisions and consider jets clustered with Cambridge-Aachen (C/A) algorithm \cite{Dokshitzer:1997in,Wobisch:1998wt} in the regime where $v\ll\zc\ll1$, where $v$ denotes the value of the observable. Although our results are derived in the context of $e^+e^-$ collisions, the NNLL results are process independent, allowing them to be used at a hadron collider, provided appropriate NLO matching is carried out. Predictions for hadron collider phenomenology will typically require results for gluon initiated jets as well as quark initiated jets, therefore phenomenological results for hadron colliders are left to future work.

We begin, in section \ref{sec:deffinitions}, by presenting the quantity to be calculated and laying out the resummation formalism we will use to do this. In section \ref{sec:multiple_emissions}, we then recap the NLL resummation of groomed observables and examine the treatment of multiple emission effects, showing that these start at N$^3$LL accuracy. The way in which we carry out the resummation to NNLL accuracy is then outlined in section \ref{sec:NNLL_Structure} in which we show how the resummation can be computed as an inclusive piece, which is evaluated in section \ref{sec:simplified}, and a clustering term, calculated in section \ref{sec:clustering}. Our resummation formula is valid in the small $\zc$ limit, nevertheless, in section \ref{sec:FiniteZ} we also include finite $\mathcal{O}(\zc)$ corrections at NLL accuracy using the results of \cite{Marzani:2017mva}. We then define our NLO matching procedure in section \ref{sec:matching} and finally present phenomenological predictions for three groomed event (jet) observables: the heavy jet mass and two angularities: the width ($\lambda^{\text{WTA}}_{\beta=1}$) and Les Houches angularity ($\lambda^{\text{WTA}}_{\beta=0.5}$) \cite{Andersen:2016qtm} in section \ref{sec:Results}. The angularities are defined as 
\begin{equation}
\lambda_{\beta}^{\mathrm{WTA}}=\frac{\sum_i E_i|\sin(\theta_i)|^{2-\beta}(1-|\cos(\theta_i)|)^{\beta-1}}{\sum_i E_i}\ ,
\end{equation}
where the sum over $i$ runs over all particles and $\theta_i$ is the angle between the particle, $i$, and the winner takes all (WTA) axis \cite{Larkoski:2014uqa}. Finally, in section \ref{sec:conclusion}, we comment on the context of our results and possible developments of this work.

\section{Observable definition and resummation formalism}\label{sec:deffinitions}

Our goal is to derive the cumulative distribution for additive jet shapes computed on jets groomed with mMDT. This is given by the number of events where the value of the observable is below $v$, namely 
\begin{align}\label{eq:cumverybasic}
	\Sigma(v;\zc) = \frac{1}{\sigma_0} \sum_{n=0}^\infty \int \sd \sigma_n\, \Theta\left(v-V^{\mathrm{mMDT}}(\tilp,k_1,...,k_n)\right) \ \ ,
\end{align}
where $\sd \sigma_n$ denotes the emission probability of producing $n$ secondary partons (including virtual corrections at all orders), $\sigma_0$ is the Born cross-section for $e^+ e^-$ collisions and $\zc$ is the mMDT parameter to be defined below. We work 
up to NNLL accuracy in $\ln(v)$, in the regime where $v\ll\zc\ll1$. Taking $v<\zc$ restricts us to the region where grooming is active. In this region the distribution is at most single logarithmic in $v$ as the argument of soft logarithms becomes $\zc$, whereas for $v>\zc$ one finds that the distribution coincides with that of the un-groomed observable, which is double logarithmic. For phenomenological purposes it is typical to take $\zc\simeq 0.1$, making resummation of $\ln(\zc)$ unnecessary. This hierarchy of scales also allows us to ignore terms suppressed by powers of $v$, $\zc$ and $v/\zc$. For simplicity we work in the context of $e^+e^-$ collisions and separate the event into two hemispheres. For concreteness, the hemispheres are separated by a plane perpendicular to the thrust axis, or equivalently, for the purpose of our calculation, a plane perpendicular to the initial $q\bar{q}$ pair. We note that the exact boundary of the hemisphere is unimportant for our purposes and it is only important that the axis used to define the hemispheres reduces to the initial $q\bar{q}$ pair in the soft and collinear limit. For example, as we will do in section \ref{sec:pheno}, the axis of the leading C/A jet in the event can be used to define the two hemispheres. We then run mMDT on each hemisphere and compute our observable on each of the groomed hemispheres\footnote{We remind the reader that even though we compute the observable on separate hemispheres, the observable is still global, and so we are not affected by non-global logarithms.}. We then require that the larger value of the observable from the two hemispheres is less than $v$. We can therefore calculate the cumulant for a single hemisphere, and take the square of it to obtain the full result:\footnote{Our formalism also applies naturally to event shapes where the contributions of the two hemispheres is summed, e.g. the thrust variable.}
\begin{equation}\label{eq:SigmaL}
\Sigma(v;\zc)=\Sigma_\ell^2(v;\zc) \ ,
\end{equation}
where $\Sigma_\ell^2(v;\zc)$ is the cumulant for a single hemisphere.

The mMDT grooming procedure starts by clustering the particles in a hemisphere using the C/A algorithm and proceeds in the following way:
\begin{enumerate}
\item Undo the last clustering in the sequence to obtain two branches, $i$ and $j$.
\item If the softer branch does not satisfy $\frac{\min(E_j,E_i)}{E_i+E_j}>\zc$ then it is discarded and the groomer returns to to step 1. 
\item If $\frac{\min(E_i,E_j)}{E_i+E_j}>\zc$ is satisfied, the groomer stops and the groomed hemisphere contains all of the particles in both $i$ and $j$.
\end{enumerate}
We define $V(\tilp,,k_1,...,k_n)$ to be a rIRC safe additive observable \cite{Banfi:2004yd}, which in the soft and collinear limit can be parametrised as

\begin{equation}\label{eq:softColObservable}
	V_{\mathrm{s.c}}(\tilp,k_1,...,k_n)=\sum_{i=1}^n d\ \bigg(\frac{k_{ti}}{Q}\bigg)^{a+b} \frac{1}{z_i^b} \ ,
\end{equation}
where $d$ is a normalisation constant that we set to unity throughout. In addition, the $k_{ti}$ are the transverse momenta of emissions with respect to the thrust axis and $z_i=2E_i/Q$ is the energy fraction of the emission.

In this work we need the groomed variant of such an observable which we write as
\begin{equation}\label{eq:mMDT_observable}
V_{\mathrm{s.c}}^{\mathrm{mMDT}}(\tilp,k_1,...,k_n)= \sum_{i=1}^n\bigg( V_{\mathrm{s.c}}(\tilp,k_i)\, \Theta^{\mathrm{mMDT}}(k_i,\tilp,k_1,...,k_{i-1},k_{i+1},...,k_n) \bigg)\ ,
\end{equation}
where $\Theta^{\mathrm{mMDT}}(k_i,\tilp,k_1,...,k_{i-1},k_{i+1},...,k_n)=1$ if the $i^{\mathrm{th}}$ emission is retained by the groomer and zero if it is removed.

In the limit  $v \ll 1$, all emissions are constrained to be soft and/or collinear with respect to the primary $q\bar{q}$ pair. We can then write the cumulative distribution in Eq.~\eqref{eq:cumverybasic} as
\begin{linenomath}
	\begin{align}\label{eq:distribution}
		\Sigma(v;\zc)= \mathcal{H}(Q) \sum_{n=0}^\infty \frac{1}{n!} \prod_{i=1}^{n}\int[\sd  k_i]\, \mathcal{M}_{\mathrm{s/c}}^2(k_1,...,k_n) \, \Theta\left(v- V^{\mathrm{mMDT}}(\tilp,k_1,...,k_n)\right) \ ,
	\end{align}
\end{linenomath}
where, $V^{\mathrm{mMDT}}(\tilp,k_1,...,k_n)$ is defined similarly to Eq. \eqref{eq:mMDT_observable} but lifting that soft and collinear approximation on $V(\tilp, k_i)$, and the phase space measure for a single massless emission reads
\begin{linenomath}
	\begin{align}
		[\sd k] = \frac{\sd^{3-2\epsilon}k}{(2\pi)^{3-2\epsilon}} \frac{1}{2E}\ \ ,
	\end{align}
\end{linenomath}
  and $\mathcal{M}_{\mathrm{s/c}}^2(k_1,...,k_n)$ is the squared matrix-element for $n$ soft and/or collinear emissions\footnote{In practice we will always replace $\mathcal{M}_{\mathrm{s/c}}^2(k_1,...,k_n)$ with a more definite approximation related to our accuracy, where at most two emissions are not fully factorised, and the rest are treated as independent.} and the factor of $1/n!$ is the symmetry factor for $n$ identical particles. Although eq. \eqref{eq:distribution} is finite in $4$ dimensions, both the real and virtual components of it are separately divergent, and hence are specified in $4-2\epsilon$ dimensions. The function $\mathcal{H}(Q)$ encodes the (normalized) all orders virtual corrections to the Born process. For two hard legs, due to trivial colour correlations, $\mathcal{H}(Q)$ takes a simple form given in \cite{Dixon:2008gr,Banfi:2018mcq}. For groomed observables at NNLL accuracy we can express $\mathcal{H}(Q)$ as 
  \begin{multline}\label{eq:virtuals_def_2}
  		\mathcal{H}(Q)=\\\left(1 + \frac{C_F\alpha_s }{2\pi} H_1\right)\, \exp\left(-\int^Q [\sd  k] \mathcal{M}_{\mathrm{soft}}^2(k)\right) \, \exp\left(-\sum_\ell \int^{Q^2}  \frac{\sd k_t^2}{k_t^{2(1+\epsilon)}} \gamma_\ell \left(\alpha_s\left(k_t^2\right)\right)  \right)\ ,
  \end{multline}
  where we have defined $\alpha_s\equiv\alpha_s(Q^2)$, which we use throughout this article. $H_1$ is a multiplicative constant obtained by matching Eq.~\eqref{eq:virtuals_def_2} onto the quark form factor at one loop \cite{Banfi:2018mcq}, and is given by $H_1=\left(\pi^2-8\right)$. Further, $\mathcal{M}_{\mathrm{soft}}^2(k)$ denotes the soft squared amplitude for a single emission, with the inclusion of the CMW (physical) coupling \cite{Catani:1990rr}. The explicit form of $\mathcal{M}_{\mathrm{soft}}^2(k)$ is given below in Eq. \eqref{eq:Msoft}. Finally, the hard collinear anomalous dimension,
  \begin{linenomath}
  	\begin{align}
  		\gamma_\ell \big(\alpha_s(k_t^2)\big)=\frac{C_F\alpha_s(k_t^2)}{2 \pi} \left(\gamma^{(0)}_{\mathrm{h.c}}+\frac{\alpha_s(k_t^2)}{2\pi}\gamma^{(1)}_{\mathrm{h.c}} \right)
  	\end{align}
  \end{linenomath}
  is needed up to $\mathcal{O}(\alpha^2_s)$ for NNLL accuracy, with the coefficients, the endpoint contributions to the DGLAP splitting kernels \cite{FURMANSKI1980437,CURCI198027}, given by:
  \begin{linenomath}
  	\begin{align}\label{eq:anomcoeffs}
  		\gamma_{\mathrm{h.c}}^{(0)} &= -\frac{3}{2} \ \ , \\
  		\gamma_{\mathrm{h.c}}^{(1)} &= - C_F \left(\frac{3}{8} -\frac{\pi^2}{2}+6 \zeta(3)  \right)
  		- C_A \left(\frac{17}{24}+\frac{11 \pi^2}{18} -3 \zeta(3) \right)
  		+ T_R n_f \left(\frac{1}{6}+\frac{2\pi^2}{9} \right)\ .
  	\end{align}
  \end{linenomath}
We introduce the Sudakov decomposition of any light-like momentum
\begin{equation}\label{eq:Sudvar}
	k^\mu= z^{(1)}  p_1^\mu + z^{(2)} p_2^\mu+\kappa^\mu \ \ ,
\end{equation}
where $\{p_1,p_2\}$ are reference vectors corresponding to the Born level $q\bar{q}$ pair. Explicitly, we have $p_1=\frac{Q}{2}(1,\vec{n}_T)$ and $p_2=\frac{Q}{2}(1,-\vec{n}_T)$, with $\vec{n}_T$ a unit vector lying along the thrust axis and $\kappa$ is a vector transverse to $p_1$ and $p_2$. The transverse momentum, with respect to the thrust axis, is then $k_t^2=-\kappa^2$. The thrust axis partitions any event into two hemispheres, each containing either of the $q \bar{q}$ pair. As noted before, the physical final state momenta of the hard system is denoted instead by $\{\tilde{p}_1,\tilde{p}_2\}$. We further note that the physical energy fraction of an emission is given by:
\begin{align}\label{eq:zzl}
	z = z^{(1)} + z^{(2)}  \ \ ,
\end{align}
 such that in the collinear limit with respect to $q$ or $(\bar{q})$, the energy fraction $z$ coincides with either $z^{(1)} $ or $z^{(2)} $, depending on the which hemisphere the emission is in.

With Eq.~\eqref{eq:Sudvar} in hand we can specify the remaining ingredients in Eq.~\eqref{eq:virtuals_def_2}. Up to NNLL accuracy we have
\begin{linenomath}
	\begin{align}\label{eq:Msoft}
		[\sd k] \mathcal{M}_{\mathrm{soft}}^2(k)= \sum_{\ell=1}^2 \frac{C_F\bar{\alpha}_s(k^2_t)}{\pi}    \frac{\sd z^{(\ell)}}{z^{(\ell)}}    \frac{\sd k_t^2}{(k_t^{2})^{1+\epsilon}}\frac{\sd \phi}{2\pi} \left(1+\frac{\bar{\alpha}_s(k_t^2)}{2\pi} K_{\mathrm{CMW}}\right) \Theta\left(z^{(\ell)}  - \frac{k_t}{Q}\right)\ ,
	\end{align}
\end{linenomath}
where the sum runs over the two hemispheres, i.e. $\ell=\{1,2\}$, and  $K_{\mathrm{CMW}}=C_A\left(\frac{67}{18}-\frac{\pi^2}{6}\right)-T_Rn_f\frac{10}{9}$ is the CMW coupling \cite{Catani:1990rr}. Essentially, this amounts to replacing the double-soft function with its inclusive limit by integrating over the branchings of the soft gluon. The lower limit on the light-cone variables $z^{(\ell)}$ delineates the hemisphere boundary
and we defined
\begin{align}
	\bar{\alpha}_s = \alpha_s \frac{ \left(4\pi \mu^2\right)^\epsilon }{\Gamma(1-\epsilon)} \ \ .
\end{align}

\section{Recap of NLL resummation}\label{sec:multiple_emissions}

Before discussing how Eq.~\eqref{eq:distribution} can be evaluated up to NNLL accuracy, we shall recap the NLL resummation of this type of observable, which is known from Ref.~\cite{Dasgupta:2013ihk}. In doing so we will pay particular attention to the effect of multiple emissions contributing to the value of the observable, showing that for the observables considered in this work such effects have an impact starting at N$^3$LL, and so can be neglected at our accuracy. 

Though we shall not attempt to derive them we will also discuss two types of terms, which despite being formally NNLL, naturally fit into an NLL resummation framework. These are terms originating from use of the CMW scheme for the coupling \cite{Catani:1990rr}, and the term usually referred to as $C_1$, which captures the $\mathcal{O}(\alpha_s)$ terms in the distribution which are not logarithmically enhanced, but survive in the limit that $v$ and $\zc$ are taken to zero.

We can evaluate $\Sigma(v;\zc)$, defined in Eq.~\eqref{eq:distribution}, to NLL accuracy by considering a sequence of independent emissions which are strongly ordered in angle and inclusive of their branchings. In this limit the squared matrix element factorises into a product of independent emissions as 
\begin{equation}
\mathcal{M}_{\mathrm{s/c}}^2(k_1,...,k_n)=\prod_{i=1}^n\mathcal{M}_{\mathrm{s/c}}^2(k_i)
\end{equation}
with the superscript $\mathrm{s/c}$ denoting that the matrix elements should either be evaluated in the soft and/or collinear approximation. At NLL accuracy, nevertheless, the collinear approximation is sufficient for mMDT groomed observables \cite{Dasgupta:2013ihk} and thus we write (in d=4 dimensions):
\begin{equation}\label{eq:collinearME}
	\int [\sd k] \mathcal{M}_{\mathrm{c}}^2(k)=\frac{C_F}{2\pi} \int_0^1 \sd z\, p_{gq}(z) \int_0^{z^2Q^2} \frac{\sd k_t^2}{k_t^2} \alpha_s(k_t^2)   \ \ ,
\end{equation}
where the argument of the coupling is set to the transverse momentum squared of the emission, and the upper limit on the $k_t^2$ integral denotes the hemisphere boundary, which at NLL accuracy is immaterial because logarithms have collinear origin.\footnote{In the current section, as well as Sects.~\ref{sec:NNLL_Structure} and \ref{sec:simplified}, we drop the azimuthal measure from our formulae.} In addition, it is sufficient to treat the transverse momentum appearing in the argument of the coupling in the soft and collinear approximation. Finally, the lowest-order splitting function reads
\begin{align}\label{eq:losplit}
	p_{gq}(z)= \frac2z - (2-z) \ \ ,
\end{align}
where we remind the reader that $z$ denotes the energy fraction of the emission.

At NLL, it also suffices to consider the observable in the soft-collinear limit as per Eq.~\eqref{eq:mMDT_observable}. Moving to the derivation, the phase space for each emission can be partitioned into a region with $z<\zc$ and a region with $z>\zc$ allowing us to write
\begin{linenomath}
	\begin{equation}\label{eq:EnergyPartitioning}
		\int[\sd  k]\mathcal{M}_{\mathrm{c}}^2(k)= \frac{C_F}{2\pi} \int_0^1 \frac{2 \sd z}{z} \int_0^{z^2Q^2}\frac{\sd  k_{t}^2}{k_{t}^2} \alpha_s\big(k_{t}^2\big) \Theta(\zc-z) + \int_0^1 \frac{\sd  v'}{v'} R_{\mathrm{NLL},\ell}'(v';\zc) \ \ ,
	\end{equation}
\end{linenomath}
where in the first term on the right hand side, which covers the phase space with $z<\zc$, we have replaced the splitting function by its soft divergent piece, as here the hard piece, i.e. $(2-z)$ in Eq.~\eqref{eq:losplit}, can only generate power corrections in $\zc$. The second term covers the phase space with $z>\zc$ and we have defined the logarithmic derivative of the Sudakov radiator\footnote{Here we effectively replace an integral over $k_t$ with a logarithmic integral over $v$ by virtue of the Dirac delta function in Eq.~\eqref{eq:RPNLL}. We also remind the reader that Eq.~\eqref{eq:RPNLL} captures only NLL terms.} 
\begin{linenomath}
	\begin{multline}\label{eq:RPNLL}
		R_{\mathrm{NLL},\ell}'(v;\zc)= \frac{C_F}{2\pi} \int_{\zc}^1 \sd z \bigg(\frac{2}{z}+\gamma^0_{\mathrm{h.c}}\delta(1-z) \bigg)\int_0^{z^2Q^2}  \frac{\sd  k_t^2}{k_t^2}v\, \delta\left(V_{\mathrm{s.c}}(z,k_t)-v\right) \ \ ,
	\end{multline}
\end{linenomath}
in which we have replaced the splitting function with $2/z+\delta(1-z)\gamma^0_{\mathrm{h.c}}$, the effect of which is to remove power corrections in $\zc$ and NNLL terms.

We can re-write the virtual corrections (Eq.~\eqref{eq:virtuals_def_2}) in the spirit of Eq.~\eqref{eq:EnergyPartitioning} by taking the collinear limit of Eq.~\eqref{eq:Msoft} and partitioning the energy fraction integral in the exponent into regions where $z>\zc$ and $z<\zc$ to write (for a single hemisphere)
\begin{multline}\label{eq:virtualsParam}
\mathcal{H}_{\mathrm{NLL},\ell}(Q)=\\ \exp\left[-\int_0^1 \frac{\sd  v'}{v'} R_{\mathrm{NLL},\ell}'(v';\zc)-\frac{C_F}{2\pi} \int_0^1 \frac{2\sd z}{z} \int_0^{z^2Q^2}\frac{\sd  k_{t}^2}{k_{t}^2} \alpha_s\big(k_{t}^2\big) \Theta(\zc-z)\right] \ ,
\end{multline}
where we have kept only the leading term in anomalous dimension, and dropped the $H_1(Q)$ and $K^{\mathrm{CMW}}$ terms, as is consistent with NLL accuracy. 

The strong ordering of emissions means that each branch of the C/A clustering sequence ,that is examined by the groomer for the $\zc$ condition, will consist of a single emission from the initial quark, inclusive of it's branchings. Furthermore, as shown in appendix \ref{app:softExponentiation}, we can neglect the contribution of any emission with $z_i<\zc$ to the observable whilst only neglecting power corrections in $v$ and $\zc$.\footnote{We stress that this does not to the apply to the NNLL clustering correction calculated in section \ref{sec:clustering}.} Physically, this is because such emissions are either groomed away, or are both softer and more collinear than the emission that dominates the value of the observable. We therefore drop any emission with $z_i<\zc$ from $V_{\mathrm{s.c}}^{\mathrm{mMDT}}(\tilp,k_1,...,k_n))$ in Eq.~\eqref{eq:distribution}. Expressing $\Theta(v-V_{\mathrm{s.c}}^{\mathrm{mMDT}}(\tilp,k_1,...,k_n))$ as its Laplace representation we can then write Eq.~\eqref{eq:distribution}, for a single hemisphere at NLL accuracy as\footnote{We are able to express real corrections in this form because we consider event shapes, which by construction do not depend on the rapidity fraction \cite{Banfi:2016zlc}.} 
\begin{linenomath}
	\begin{multline}\label{eq:MultipleEmission}
		\Sigma_{\mathrm{NLL},\ell}(v;\zc)= \exp\left[-\int_0^1 \frac{\sd  v'}{v'} R_{\mathrm{NLL},\ell}'(v';\zc)-\frac{C_F}{2\pi} \int_0^1 \frac{2\sd z}{z} \int_0^{z^2Q^2}\frac{\sd  k_{t}^2}{k_{t}^2} \alpha_s\big(k_{t}^2\big) \Theta(\zc-z)\right] \\ \int_c \frac{\sd  \nu}{2\pi i \nu}e^{-\nu v} \sum_{n=0}^\infty \frac{1}{n!}  \bigg( \int_0^1 \frac{\sd  v'}{v'} e^{\nu v'}R_{\mathrm{NLL},\ell}'(v';\zc)\\ + \frac{C_F}{2\pi} \int_0^1 \frac{2 \sd z}{z} \int_0^{z^2Q^2}\frac{\sd  k_{t}^2}{k_{t}^2} \alpha_s\big(k_{t}^2\big) \Theta(\zc-z) \bigg)^n\ ,
	\end{multline}
\end{linenomath}
where the first line is the virtual term, and the second line represents real corrections. Note that in the above expression there is no factor of $e^{-\nu v'}$ in the term accounting for real emissions with $z<\zc$, as real emissions have been dropped from the observable in this region of phase space. Eq.~\eqref{eq:MultipleEmission} can then be evaluated using what are now standard techniques \cite{CATANI1991368} to give
\begin{equation}
\Sigma_{\mathrm{NLL},\ell}(v;\zc)= \frac{\exp\left[-R_{\mathrm{NLL},\ell}(v;\zc)-\gamma_{E}R_{\mathrm{NLL},\ell}'(v;\zc)\right]}{\Gamma[1+R_{\mathrm{NLL},\ell}'(v;\zc)]}\ ,
\end{equation}
where the Sudakov radiator is $R_{\mathrm{NLL,\ell}}(v;\zc)=\int_v^1 \frac{\sd v'}{v'} R'_{\mathrm{NLL,\ell}}(v';\zc)$. Here we can see that the emissions which are softer than $\zc$ have cancelled completely against the corresponding virtual corrections. Since $R_{\mathrm{NLL},\ell}(v;\zc)$ is single logarithmic, we see that expanding $\frac{\exp[-\gamma_{E}R'(v)]}{\Gamma[1+R'(v)]}$ contributes at the N$^3$LL  level and thus can be neglected.\footnote{This may not be the case for non-additive observables such as the broadening \cite{Rakow:1981qn}.}

In the above derivation we have neglected terms proportional to $K_{\mathrm{CMW}}$ which are NNLL for groomed observables and do emerge in the full NNLL calculations of Ref.~\cite{Anderle:2020mxj}. However, these terms are naturally part of a strongly ordered NLL resummation and can be included by a modification to the radiator \cite{Dasgupta:2013ihk} 
\begin{multline}\label{eq:Sudprime}
	R_{\mathrm{NLL}^\prime,\ell}(v;\zc) = \frac{C_F}{2\pi} \int_0^1 \sd  z \int_0^{z^2Q^2} \frac{\sd  k_t^2}{k_t^2} \alpha_s(k_t^2) \bigg[\frac{2}{z}\bigg(1+ \frac{\alpha_s(k_t^2)}{2\pi}K_{\mathrm{CMW}}\bigg)\\ +  \int_0^1 \sd z \int_0^{z^2 Q^2} \frac{\sd k_t^2}{k_t^2}\, \alpha_s(k_t^2) \gamma^{(0)}_{\mathrm{h.c}} \delta(1-z) \bigg]\Theta\left(V_{\mathrm{s.c}}(z,k_t)-v\right)\Theta(z-\zc)\ .
\end{multline}
where the primed notation denotes that NNLL effects which fit into the strongly ordered framework are included. The other NNLL term which naturally fits into a strongly ordered resummation is the the leading order constant, which on physical grounds must factorise from the exponential (Sudakov factor). This contribution can naturally be accommodated within a strongly ordered resummation through a coefficient, denoted by $C_1$. Translating to the full result for both hemispheres, as per Eq. \eqref{eq:SigmaL}, we can then write the NLL result, supplemented with the aforementioned NNLL terms as
\begin{equation}\label{eq:LL}
\Sigma_{\mathrm{NLL}^\prime}(v;\zc)=\left(1+\frac{\alpha_s C_F}{2\pi}C_1\right) \exp[-R_{\mathrm{NLL}^\prime}(v;\zc)] \ ,
\end{equation}
where the full radiator is related to the radiator for a single hemisphere simply by a factor of two: $R_{\mathrm{NLL'}}(v;\zc)=2R_{\mathrm{NLL'},\ell}(v;\zc)$.

In the next section, we will see that, along with the addition of a term to account for the effect of the C/A clustering sequence, the remarkably simple structure of Eq.~\eqref{eq:LL} persists at NNLL accuracy, with remaining NNLL terms accounted for through terms related to the standard $C_1$ by running coupling effects, and by evaluating the hard collinear part of $R(v;\zc)$ up to NNLL accuracy \cite{Banfi:2018mcq}. 

\section{Structure of NNLL resummation}\label{sec:NNLL_Structure}

\begin{figure}[h]
\centering
\includegraphics[width=0.6\textwidth]{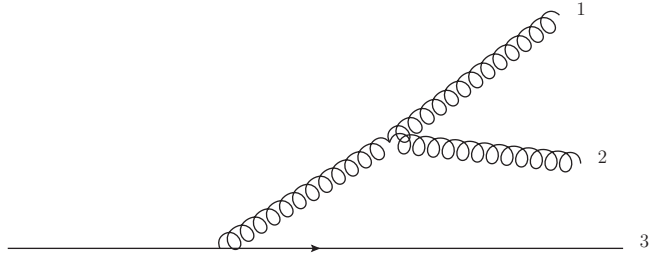}
\caption{A diagram showing a quark emitting a gluon, which then decays to a pair of gluons. Using our simplified groomer, the two gluons are always clustered together first, the resulting branch is then clustered with the quark.}
\label{fig:Simplified_groomer}
\end{figure}

The groomed jet mass distribution was investigated, in the relevant triple-collinear limit at order $\alpha_s^2$, in Refs.~\cite{Anderle:2020mxj}. In addition, the distribution of a generic observable has been considered in Ref.~\cite{Dasgupta:2021hbh}. It was demonstrated that the remaining part of the NNLL groomed jet mass result, which is not related to the strongly-ordered picture (i.e. not already included within Eq.~\eqref{eq:LL}), is structured as an inclusive hard-collinear piece, often referred to as $B^{(2)}$ \cite{COLLINS1981381,COLLINS1982446,KODAIRA198266,KODAIRA1983335}, plus a term accounting for the effect of the C/A clustering sequence in the grooming procedure. This motivates us to structure the resummation in a similar way, by considering a suitably inclusive version of the groomed observable, which is added to a clustering correction to give the resummed distribution we seek. This organization of our resummed predictions is quite natural, e.g. see Ref.~\cite{Banfi:2012jm}. Motivated by this we define an observable, $V^{\mathrm{simp.}}(\{\tilde{p}\},k_1,...,k_n)$, computed using a \textit{simplified groomer} which functions exactly as mMDT except that the C/A clustering sequence is replaced by one where partons originating from a common parent are clustered together. For example, in Fig.~\ref{fig:Simplified_groomer}, the gluons (1 and 2) are first clustered together, irrespective of which pairwise angle between any of the three partons is smallest, followed by clustering the resulting branch with the quark.

The integrated distribution (Eq.~\eqref{eq:distribution}) is then written as
\begin{align}\label{eq:NNLLpartition}
  \Sigma(v;\zc) = \Sigma^{\text{simp.}}(v;\zc) + \Sigma^{\text{clust.}}(v;\zc) \ ,
\end{align}
 where the first contribution reads
\begin{multline}\label{eq:simpcum}
  \Sigma^\text{simp.}(v;\zc)=\\ \mathcal{H}(Q) \sum_{n=0}^\infty \frac{1}{n!} \int \prod_{i=1}^{n}[\sd  k_i]\mathcal{M}_{\mathrm{s/c}}^2(k_1,...,k_n) \Theta\left( v - V^{\mathrm{simp.}}(\tilp,k_1,...,k_n)\right) \ \ ,
\end{multline}
and represents the integrated distribution computed using the inclusive groomer. The second contribution embodies the difference between the actual and simplified groomers as follows
\begin{multline}\label{eq:clustcum}
\Sigma^{\text{clust.}}(v;\zc)= \mathcal{H}(Q) \sum_{n=0}^\infty \frac{1}{n!} \prod_{i=1}^{n}\int[\sd  k_i]\mathcal{M}_{\mathrm{s/c}}^2(k_1,...,k_n)\\
\times \big[ \Theta\left(v - V^{\mathrm{mMDT}}(\tilp,k_1,...,k_n)\right)-\Theta\left(v - V^{\mathrm{simp.}}(\tilp,k_1,...,k_n)\right)\big] \ \ .
\end{multline}
The clustering correction starts contributing only at NNLL, and arises from the regions of phase space where the C/A clustering sequence differs from that of our simplified groomer. This leads to different sets of emissions being groomed away by the two groomers \cite{Anderle:2020mxj}. Sections \ref{sec:simplified} and \ref{sec:clustering} describe in detail the NNLL resummation of the pieces in Eqs.~\eqref{eq:simpcum} and \eqref{eq:clustcum} respectively.

\section{The simplified groomer at NNLL accuracy}\label{sec:simplified}

In this section we compute the NNLL resummed cumulant for the simplified groomer, as given in Eq.~\eqref{eq:simpcum}. From Refs.~\cite{Anderle:2020mxj} we can see that $\Sigma^{\mathrm{simp.}}(v;\zc)$ should contain all of the structures given in Eq.~\eqref{eq:LL}; the ones which stem from strongly ordered configurations of emissions, as well as hard-collinear terms which first enter at $\mathcal{O}(\alpha_s^2)$, proportional to $B^{(2)}$ \cite{COLLINS1981381,COLLINS1982446,KODAIRA198266,KODAIRA1983335}.

We have already presented an NLL$^\prime$ result in Sect.~\ref{sec:multiple_emissions} and shown that multiple emission effects start at N$^3$LL, and so can be neglected at the NNLL accuracy we ultimately obtain. We will now show, in Sect.~\ref{sec:hardCollinear}, how this NLL$^\prime$ result can be modified to include the NNLL hard collinear terms, completing the NNLL evaluation of $\Sigma^{\mathrm{simp.}}(v;\zc)$. We will uncover the almost exact similarity between the groomed simplified cumulant, $\Sigma^{\mathrm{simp.}}(v;\zc)$, and the ungroomed version of the observable. In particular, the structure in Eq.~\eqref{eq:NNLLpartition} is chosen to make contact with already known results for un-groomed observables in the \texttt{ARES} formalism \cite{Banfi:2014sua,Banfi:2018mcq}. An example calculation of the observable dependent hard-collinear terms is provided for the angularities with respect to the WTA axis in Sect.~\ref{sec:C1}.

\subsection{NNLL hard collinear terms: \texorpdfstring{$B^{(2)}$}{B2}} \label{sec:hardCollinear}

In this section we show how the NNLL hard-collinear term, $B^{(2)}$, arises in our resummation. At $\mathcal{O}(\alpha_s^2)$, for a general observable that satisfies Eq.~\eqref{eq:softColObservable}, the term proportional to $B^{(2)}$ is of the form 
\begin{equation}
  -B^{(2)} \int_{v^{\frac{2}{a+b}}Q^2}^{Q^2}\left(\frac{\alpha_s\big(k_t^2\big)}{2\pi}\right)^2\frac{\sd  k_t^2}{k_t^2}
\end{equation}
and it is known that \cite{deFlorian:2000pr,Catani:2000vq,deFlorian:2004mp}
\begin{equation}\label{eq:B2} 
B^{(2)}=C_F\gamma^{(1)}_{\mathrm{h.c}}+C_F b_0 X_v \ \ ,
\end{equation}
where $\gamma^{(1)}_{\mathrm{h.c}}$ is given in Eq.~\eqref{eq:anomcoeffs}, $b_0=(11C_A-4T_Rn_f)/6$ and $X_v$ is an observable dependent coefficient. This term was identified, for the groomed jet mass, via an NLO calculation in \cite{Anderle:2020mxj}, which was found to be $X_\rho=\frac{\pi^2}{3}-\frac72$, for a single hemisphere. The observation was made that this is the same as for the un-groomed jet mass \cite{Banfi:2014sua,Banfi:2018mcq,Chien:2010kc}, as one might expect since hard emissions are not affected by grooming. We note that at our prescribed level of accuracy there is some freedom in how these terms are included in the resummation, i.e. whether or not they are exponentiated.

The $\gamma^{(1)}_{\mathrm{h.c}}$ term is universal for all observables and is just part of the quark form factor \cite{Banfi:2018mcq,deFlorian:2001zd,deFlorian:2000pr}. It should therefore sit in the  Sudakov radiator as it does for un-groomed observables \cite{Banfi:2018mcq}, which then reads (for a single hemisphere)
\begin{linenomath}
	\begin{multline}\label{eq:Sudakov1}
		R_{\text{NNLL},\ell}(v;\zc)=  \frac{C_F}{2\pi}\times \\ \int_0^1 \frac{2\sd z^{(\ell)}}{z^{(\ell)}} \int_0^{\left(z^{(\ell)} Q\right)^2} \frac{\sd  k_t^2}{k_t^2} \alpha_s(k_t^2) \bigg(1+ \frac{\alpha_s(k_t^2)}{2\pi}K_{\mathrm{CMW}}\bigg)     \Theta\left(V_{\mathrm{s.c}}(z^{(\ell)},k_t)-v\right)   \Theta(z-\zc)+\frac{C_F}{2\pi} \times\\ \int_0^1 \sd z \int_0^{z^2 Q^2} \frac{\sd k_t^2}{k_t^2}\, \alpha_s(k_t^2) \left(\gamma^{(0)}_{\mathrm{h.c}}+\frac{\alpha_s(k_t^2)}{2\pi}\gamma^{(1)}_{\mathrm{h.c}}\right)\delta(1-z) \Theta\left(V_{\mathrm{s.c}}(z,k_t)-v\right)\Theta(z-\zc)\ .
	\end{multline}
\end{linenomath}
We note importantly that at NNLL the soft limit, i.e. the first contribution in Eq.~\eqref{eq:Sudakov1}, must be employed exactly. In particular, it is not sufficient to enforce the collinear limit as we did at the NLL level in Eq.~\eqref{eq:Sudprime}. In the soft wide-angle region, there is a mismatch between the energy fraction of the emission, $z = z^{(\ell)} + k_t^2/\left(z^{(\ell)}Q^2\right)$, and the light-cone momentum fraction $z^{(\ell)}$ which results in a constant at $\mathcal{O}(\alpha_s)$, i.e. NNLL contribution, which here appears in the Sudakov factor. Within NNLL accuracy we are free to, and indeed do, remove this term from the Sudakov factor and instead include it in the $C_1$ term. This leaves the Sudakov radiator free of soft wide angle terms, and thus process independent. The evaluation of Eq.~\eqref{eq:Sudakov1} is straightforward and is given in Appendix.~\ref{app:NNLLSud}. When evaluating $R_{\text{NNLL}}(v;\zc)$, the terms which start at $\mathcal{O}(\alpha_s)$, should have the coupling evaluated at two loops in order to capture a set of NNLL terms, whilst for the terms which start at $\mathcal{O}(\alpha_s^2)$, one loop accuracy is sufficient. 

We now discus how the the observable dependent piece of the hard-collinear terms proportional to $b_0 X_v$ are included in the resummation. As already stressed, the hard-collinear pieces of the resummation are the same as for un-groomed observables and so can be resummed using the same methods. By examining Refs.~\cite{Banfi:2014sua,Banfi:2018mcq,Dasgupta:2021hbh}, we can see that the observable dependent part of $B^{(2)}$, the $b_0 X_v$ term, is in fact related to the leading-order result and can be found by computing the $C_1$ term of eq. \eqref{eq:LL} with a running coupling with argument set to $k_t^2$, as we shall further elaborate on below. Within the ARES formalism \cite{Banfi:2014sua,Banfi:2018mcq} this can be seen by examining the functions $C^1_{\mathrm{h.c}}$ and $\delta\mathcal{F}_{\mathrm{rec}}$, observing that they reproduce our $b_0 X_v$ term and the corresponding parts of $C_1$.

The $C_1$ term of Eq.~\eqref{eq:LL} was calculated for the heavy hemisphere mass in Ref.~\cite{Anderle:2020mxj} ({\em cf.} section 3 therein), and reads $C_1=-2 \ln 2 \left(- 4 \ln \zc-3\right) -1$, if the jet mass is normalized to $(Q/2)^2$. So as to reduce the number of terms in formulae we instead adopt the normalization $\rho=M^2_{\mathrm{H}}/Q^2$ as this removes the factors of $\ln(2)$ and thus $C_1=-1$. If we absorb the factor of $\frac{C_F\alpha_s}{2\pi}$ into the definition of $C_1$ and re-calculate  it with the argument of the coupling set to $k_t^2$,\footnote{In the argument of the coupling we have used $k_t$ in the soft and collinear limit, i.e. $k_t=z\theta Q/2$, although any definition of $k_t$ which coincides with this in the soft limit would be equivalent within NNLL accuracy.} which we denote by $C^{\mathrm{r.c}}(v)$ we find 
\begin{multline}\label{eq:CrcRho}
C^{\mathrm{r.c}}(\rho)=\frac{C_F\alpha_s}{2\pi}\bigg(\frac{1}{1-\lambda_\rho}\bigg(7-\frac{2\pi^2}{3}\bigg)+\frac{1}{1+2\lambda_{\zc}}\bigg(\frac{-\pi^2}{3} \bigg) +\left(\pi^2-8\right)\bigg)\\
\simeq-\frac{C_F\alpha_s}{2\pi}+C_F\bigg(\frac{\alpha_s }{2\pi}\bigg)^2b_0 \left(\frac{2\pi^2}{3}-7\right)\ln\rho +\mathcal{O}(\alpha_s^3)+\mathrm{N}^3\mathrm{LL}\ ,
\end{multline}
where we have defined $\lambda_x=\beta_0\alpha_s\ln1/x$ and $b_0 = 2\pi \beta_0$. From the second line of Eq.~\eqref{eq:CrcRho} we see that this exactly reproduces both the $C_1$ and $b_0 X_\rho=b_0\left(\frac{2\pi^2}{3}-7\right)$ terms identified in \cite{Anderle:2020mxj,Dasgupta:2021hbh}. The details of how $C^{\mathrm{r.c}}(v)$ is calculated are given in Sect.~\ref{sec:C1} where we illustrate the calculation using the example of the angularities with respect to the WTA axis. Use of the running coupling also results in terms beyond $\mathcal{O}(\alpha_s)$ which are not enhanced by logarithms of the observable but by logarithms of $\zc$, which are N$^3$LL and so can be neglected, as we have done in the expansion in the second line of Eq.~\eqref{eq:CrcRho}.

Our final NNLL resummed result for the simplified groomer now reads 
\begin{equation}\label{eq:Simplified_Result}
\Sigma_{\mathrm{NNLL}}^{\mathrm{simp}}(v;\zc)=\big(1+C^{\mathrm{r.c}}(v)\big) \exp\left[-\bar{R}_{\text{NNLL}}(v;\zc)\right] \ ,
\end{equation}
where $\bar{R}_{\text{NNLL}}(v;\zc) \equiv 2 \bar{R}_{\text{NNLL},\ell}(v;\zc)$, given in Eq.~\eqref{eq:nnllbarsud}. This result bears a strong resemblance to the NLL result of \cite{Dasgupta:2013ihk}, which was re-derived in section \ref{sec:multiple_emissions}. The pure NLL result is recovered if one neglects the $C^{\mathrm{r.c}}(v)$ term as well as the $\gamma^{(1)}_{\mathrm{h.c}}$ and $K_{\mathrm{CMW}}$ terms in the Sudakov factor.

\subsection{Calculating \texorpdfstring{$C^{\mathrm{r.c}}(v)$}{Crc(v)}}\label{sec:C1}

We now turn our attention to illustrating more precisely what is captured by the function $C^{\mathrm{r.c}}(v)$ and how it can be calculated. The computation follows the \texttt{ARES} formalism \cite{Banfi:2014sua,Banfi:2018mcq}, but tailored to groomed observables. $C^{\mathrm{r.c}}(v)$ captures the difference between the leading-order result, computed with the argument of the coupling set to $k_t^2$,\footnote{As shall be illustrated below, setting the argument of the coupling to $k_t^2$ will turn out only to be important in the collinear limit, and even then, the $k_t$ which appears in the argument of the coupling need only be correct in the soft and collinear limit.} and the part of the $\mathcal{O}(\alpha_s)$ result that is captured by the Sudakov factor. Terms suppressed by powers of $v$ or $\zc$ are neglected throughout. 

When the emission is retained by the groomer, we calculate the difference between the real contribution to the leading-order integrated distribution and the corresponding contribution that is exponentiated, which we denote by $C_{\text{col.},\ell}^{\mathrm{r.c}}(v)$. We then add to this $H_1(Q)=\frac{C_F \alpha_s}{2 \pi}(\pi^2-8)$, the piece of the leading-order virtual corrections that is not included in the Sudakov factor. In addition, as part of the definition of $C^{\mathrm{r.c}}(v)$ we include the constant term proportional to $\pi^2$ that appears in the Sudakov radiator in Eq.~\eqref{eq:g2}. The choice to expand this term in the radiator is made to ensure that the radiator is process independent.

The general expression then reads (for a single hemisphere)
\begin{align}\label{eq:C1}
C_\ell^{\mathrm{r.c}}(v) =\frac{C_F\alpha_s}{2\pi} \left(\frac{\pi^2}{2} - 4 \right)- \frac{C_F\alpha_s}{2\pi} \left(\frac{\pi^2}{6}\right) + C_{\text{col.},\ell}^{\mathrm{r.c}}(v)  \ \ ,
\end{align}
where, the first term is half of $H_1(Q)$, the second is the constant, soft-wide-angle term that we expanded from the Sudakov factor and
\begin{multline}\label{eq:C1col}
	C_{\text{col.},\ell}^{\mathrm{r.c}}(v) = \int[\sd k]\mathcal{M}^2(k;\epsilon)\Theta\left(v-V^{\mathrm{simp.}}(\tilp,k)\right) \Theta(z - \zc) \\
	-\frac{C_F}{2\pi}\int_{\zc}^1 \sd z \bigg( \frac{2}{z} - \gamma^{(0)}_{\mathrm{h.c}}\delta(1-z)\bigg) \int_{0}^{z^2Q^2} \sd z \frac{\sd  k_t^2}{k_t^{2(1+\epsilon)}} \alpha_s\big(k_t^2\big) \Theta\left(v-V^{\mathrm{simp.}}_{\mathrm{s.c}}(z,k_t)\right) \  \ ,
\end{multline}
where $\mathcal{M}^2(k;\epsilon)$ is the tree-level matrix element squared for the emission of a collinear gluon from a $q\bar{q}$ pair, retaining the full $\epsilon$ dependence. This calculation is carried out in $d=4-2\epsilon$ dimensions as the two integrals are separately divergent, however the full result is finite in four dimensions. We note that apart from the need to capture the $\epsilon$ dependent part of $\mathcal{M}^2(k;\epsilon)$, this calculation could be carried out directly in $d=4$ dimensions by combining the integrals in the first and second line. On the first line of Eq.~\eqref{eq:C1col} it is important to treat the observable correctly in the hard-collinear region of phase space rather than using the soft and collinear parametrisation of the observable as is done in the second line of Eq.~\eqref{eq:C1col} \cite{Banfi:2014sua,Banfi:2018mcq}. 

As well as correcting the observable, Eq.~\eqref{eq:C1col} also corrects the matrix element in the hard-collinear limit, which amounts to capturing terms generated by the interplay of the $\mathcal{O}(\epsilon)$ part of the splitting function and the collinear pole. As the emission is retained by the groomer ($z>\zc$), $C_{\mathrm{col.},\ell}^{\mathrm{r.c}}(v)$ depends on the precise behaviour of the observable in the hard-collinear limit, therefore, Eq.~\eqref{eq:C1col} must be evaluated on an observable-by-observable basis. Below we illustrate how $C_{\mathrm{col.},\ell}^{\mathrm{r.c}}(v)$ can be evaluated, using the angularities \cite{Larkoski:2014pca,Berger:2003iw} with respect to the WTA axis \cite{Larkoski:2014uqa} as an example.
For $e^+e^-$ colliders the angularities are defined as \cite{Berger:2003iw,Banfi:2018mcq}
\begin{equation}
\lambda_\beta^{\mathrm{WTA}}=\frac{\sum_i E_i|\sin(\theta_i)|^{2-\beta}(1-|\cos(\theta_i)|)^{\beta-1}}{\sum_i E_i}\ ,
\end{equation}
where the sum runs over all particles in the hemisphere after grooming and $\theta_i$ is the angle between a particle and the WTA axis. In the soft and collinear approximation these observables can be parametrised as per Eq.~\eqref{eq:softColObservable} with $a=1$ and $b=\beta-1$.

In the presence of a single hard-collinear emission, the angularities with respect to the WTA axis reads \cite{Banfi:2018mcq}
\begin{equation}
\lambda_\beta^{\mathrm{WTA}}=\frac{\min(z,1-z)}{(z(1-z))^\beta}\bigg(\frac{k_t}{Q}\bigg)^\beta \ \ ,
\end{equation}
from which the soft-collinear limit, $z\to 0$, is easily identified, viz.
\begin{align}
	\lambda_\beta^{\mathrm{WTA,soft}}= z^{1-\beta} \bigg(\frac{k_t}{Q}\bigg)^\beta \ \ .
\end{align}
Evaluating Eq.~\eqref{eq:C1col} is equivalent to computing the functions $C_{\mathrm{h.c}}^1$ and $\delta\mathcal{F}^{\mathrm{rec}}$ in \cite{Banfi:2014sua,Banfi:2018mcq}. With $z>\zc$, we can take the collinear approximation to $\mathcal{M}^2(k;\epsilon)$ to write 
\begin{multline}\label{eq:c1_col_0}
C_{\text{col.},\ell}^{\mathrm{r.c}}(v)= \frac{C_F}{2\pi}  \int_{\zc}^1 \sd z \, p_{gq}(z;\epsilon) \int_0^{z^2 Q^2} \frac{\sd  k_t^2}{k_t^{2(1+\epsilon)}} \alpha_s\big(k_t^2\big)  \Theta\left(\lambda_\beta^{\mathrm{WTA}}-\frac{\min(z,1-z)}{(z(1-z))^\beta}\bigg(\frac{k_t}{Q}\bigg)^\beta\right) \\
 -\frac{C_F}{2\pi}\int_{\zc}^1 \sd z \bigg( \frac{2}{z} - \gamma^{(0)}_{\mathrm{h.c}}\delta(1-z)\bigg) \int_{0}^{z^2Q^2} \frac{\sd  k_t^2}{k_t^{2(1+\epsilon)}} \alpha_s\big(k_t^2\big) 
 \Theta\left(\lambda_\beta^{\mathrm{WTA}}- z ^{1 -\beta}\bigg(\frac{k_t}{Q}\bigg)^\beta\right) \ \ ,
\end{multline}
where $p_{gq}(z;\epsilon)=\left(\frac{1+(1-z)^2}{z}+\epsilon z\right)$ is the $d$-dimensional splitting function. Eq. \eqref{eq:c1_col_0} corrects the treatment of the observable in the hard-collinear region as well as capturing terms due to the interplay of the $\mathcal{O}(\epsilon)$ term in the splitting function and the collinear pole. To evaluate this within NNLL accuracy, we can make the approximation \cite{Banfi:2014sua} that
\begin{equation}\label{eq:C1_col_start}
\alpha_s\big(k_t^2\big)\simeq\alpha_s\big((\lambda^{\mathrm{WTA}}_\beta)^{\frac{2}{\beta}}Q^2\big)=\frac{\alpha_s}{1-(2/\beta)\lambda_{\lambda^{\mathrm{WTA}}_{\beta}}} \ ,
\end{equation}
where $\lambda_x$ was defined in the previous section. With this in hand we can now  evaluate Eq.~\eqref{eq:c1_col_0} to find
\begin{equation}\label{eq:C1_col}
C_{\text{col.},\ell}^{\mathrm{r.c}}(v)=\frac{C_F}{2\pi} \frac{\alpha_s}{\big(1-(2/\beta)\lambda_{\lambda^{\mathrm{WTA}}_{\beta}}\big)}\left(-\frac{3}{\beta }+\frac{ \pi ^2}{3 \beta }-\frac{3 \ln(2)}{\beta }-\frac{2 \pi ^2}{3}+\frac{13}{2} \right)\  .
\end{equation}
Notice in particular that this computation has a smooth limit as $\zc \to 0$, and thus one could set $\zc$ to zero from the outset. 
Plugging this result back into Eq.~\eqref{eq:C1}, we can then write the expression for $C^{\mathrm{r.c}}(v)$ for the angularities with respect to the WTA axis:
\begin{equation}
C_\ell^{\mathrm{r.c}}(v)= \frac{C_F\alpha_s}{2\pi}\bigg(\frac{1}{1-(2/\beta)\lambda_{\lambda^{\mathrm{WTA}}_{\beta}}}\bigg(-\frac{3}{\beta }+\frac{ \pi ^2}{3 \beta }-\frac{3 \ln(2)}{\beta }-\frac{2 \pi ^2}{3}+\frac{13}{2}\bigg) +\bigg(\frac{\pi^2}{3}-4\bigg) \bigg)\ .
\end{equation}
Expanding this to $\mathcal{O}(\alpha_s^2)$ one finds
\begin{multline}\label{eq:crc_expansion}
C^{\mathrm{r.c}}_\ell(v)=\frac{C_F \alpha_s}{2\pi}\bigg(\frac52-\frac{3}{\beta }+\frac{ \pi ^2}{3}\left(\frac{1}{\beta}-1\right)-\frac{3 \ln(2)}{\beta } \bigg)- \\ \left(\frac{\alpha_s}{2\pi}\right)^2C_Fb_0\left(-\frac{3}{\beta }+\frac{\pi ^2}{3 \beta }-\frac{3 \ln(2)}{\beta }-\frac{2 \pi ^2}{3}+\frac{13}{2}\right)\frac{2}{\beta}\ln(\lambda^{\mathrm{WTA}}_\beta) +\mathcal{O}(\alpha_s^3)
\end{multline}
which gives the standard $C_1$ term at order $\alpha_s$ (for a single hemisphere) and a term proportional to $b_0$ at order $\alpha_s^2$, which forms part of the hard-collinear coefficient $B^{(2)}= \gamma^{(1)}_{\mathrm{h.c}}+b_0X$ where for the angularities $X_{\lambda_\beta^{\mathrm{WTA}}}$ is the coefficient of $b_0$ in Eq.~\eqref{eq:crc_expansion}.

\subsection{Relationship to resummation of un-groomed observables}\label{sec:discussion}
We now discuss how the above result for $\Sigma_{\mathrm{simp}}(v;\zc)$ is related to NNLL resummations of un-groomed observables as carried out using \texttt{ARES} \cite{Banfi:2014sua,Banfi:2018mcq}, highlighting why a number of the effects included there are relevant only beyond NNLL accuracy for groomed observables. The Sudakov factor reported in Eq.~\eqref{eq:Sudakov1} has a similar structure as that given in \cite{Banfi:2018mcq}, the only differences being the boundary of the energy fraction integrals and that here there is no $K^{(2)}$ term in the soft physical coupling as in Eq.~(3.9) of Ref. \cite{Banfi:2018mcq} as this would be N$^3$LL. Our $C_{\mathrm{col.}}^{\mathrm{r.c}}(v)$ function is identical to the terms $\delta\mathcal{F}^{\mathrm{rec}}$ and $C_{\mathrm{h.c}}^1$ in \cite{Banfi:2014sua,Banfi:2018mcq}. The \texttt{ARES} function $\delta\mathcal{F}_{\mathrm{wa}}$ would at most contribute power corrections in $\zc$ because when an emission is retained by the groomer the angle is constrained to be of the order $v/\zc \ll 1$.

The \texttt{ARES} terms $\mathcal{F}_{\mathrm{NLL}}$, $\delta\mathcal{F}_{\mathrm{h.c}}$ and $\delta\mathcal{F}_{\mathrm{s.c}}$ are related to multiple emission effects which we have shown start at N$^3$LL for the observables considered in this work, although they could come into play at NNLL for non-additive observables such as the broadening \cite{RAKOW198163}. Finally, for un-groomed observables one should, at NNLL accuracy, correct for the inclusive treatment of correlated emissions for a single  correlated pair, giving a correction starting at $\alpha_s^2\ln(v)$ which in \texttt{ARES} is called $\delta\mathcal{F}_{\mathrm{correl.}}$. This logarithm is of soft origin and for groomed observables is replaced by a logarithm of $\zc$, making the correction N$^3$LL.

\section{The clustering correction}\label{sec:clustering}

We now turn our attention to the clustering correction which is given by 
\begin{multline}\label{eq:ClusteringCorDef}
\Sigma^{\text{clust.}}(v;\zc)=\mathcal{H}(Q) \sum_{n=0}^\infty \frac{1}{n!}\prod_{i=1}^{n}\int[\sd  k_i] \mathcal{M}_{\mathrm{s/c}}^2(k_1,...,k_n)\times \\ [\Theta(v - V^{\mathrm{mMDT}}(\tilp,k_1,...,k_n))-\Theta(v - V^{\mathrm{simp.}}_{\mathrm{s.c}}(\tilp,k_1,...,k_n))],
\end{multline}
which is added to Eq.~\eqref{eq:Simplified_Result} to give $\Sigma(v;\zc)$ as in Eq.~\eqref{eq:NNLLpartition}. It is in this part of the calculation where the lack of rIRC safety of groomed observables plays a role. This is because, as we shall see, the clustering correction is generated as a result of the observable's scaling with the momentum of one emission depending on the momentum of another.

\subsection{Independent emission clustering correction}\label{sec:IndependentClustering}

\begin{figure}
\centering
\includegraphics[width=0.45\textwidth]{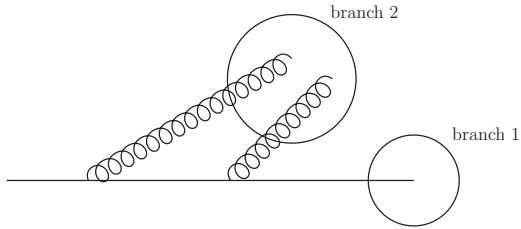}
\caption{The configuration responsible for the independent emission clustering correction, where two soft emissions with $z_i<\zc$ are de-clustered from the parent parton as a single branch which as a whole passes the $\zc$ condition.}
\label{fig:CFCFClusterDiagram}
\end{figure}

As there are some differences between the clustering corrections for independent and correlated emissions, we will first compute $\Sigma^{\text{clust.}}(v;\zc)$ considering only independent emissions. The calculation is carried out considering emissions in a single hemisphere, with a factor of two provided to account for the opposite hemisphere. This piece accounts for the fact that gluons softer than $\zc$, which in the previous section were always treated as being groomed away, can be retained by the groomer, when, due to the C/A clustering sequence, they are examined for the $\zc$ condition as part of a branch containing another independent soft gluon such that for two gluons labelled $\alpha$ and $\beta$,
\begin{equation}\label{eq:encond_ind}
 z_\alpha,z_\beta<\zc \ \ \text{and} \ \ z_\alpha+z_\beta>\zc\ ,
\end{equation}
resulting in the branch being retained by the groomer. This is exactly the clustering correction calculated for the heavy hemisphere mass in Ref. \cite{Anderle:2020mxj}. At NNLL accuracy, it is sufficient to consider that such a branch only contains two independent emissions as is depicted in Fig.~\ref{fig:CFCFClusterDiagram}. This is because requiring an extra emission in a branch results in an extra power of the coupling but no additional logarithms so that such configurations are at least N$^3$LL.

\begin{figure}
\centering
\includegraphics[width=0.45\textwidth,trim=100 550 360 60, clip]{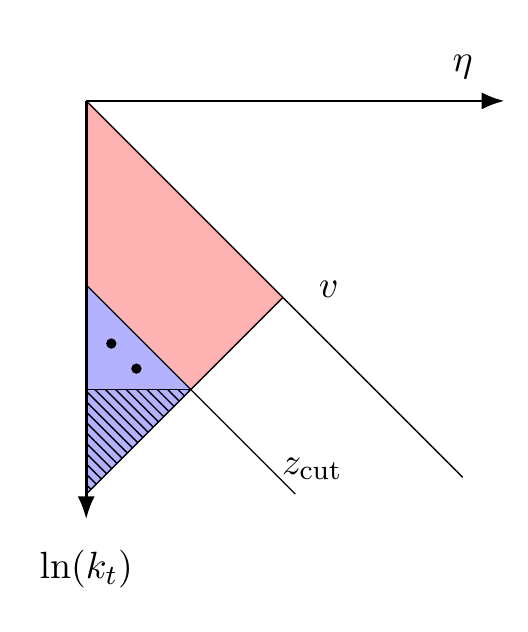}
\caption{Lund diagram showing the region of phase space responsible for the independent emission clustering correction in blue. The region shown in red is the vetoed phase space responsible for the Sudakov factor. The two dots represent possible locations in the phase space for the pair of independent emissions which generate the clustering correction. Discussion of the hashed area, which contributes in principle to the clustering correction but turns out only to beyond NNLL accuracy, is postponed to later in this section.}
\label{fig:abelianClusteringPlane}
\end{figure}

As per Eq.~\eqref{eq:ClusteringCorDef}, the clustering correction $\Sigma^{\text{clust.}}(v;\zc)$ is proportional to
\begin{equation}\label{eq:ClusteringThetas}
\Theta(v-V^{\mathrm{mMDT}}(\tilp,k_1,...,k_n))- \Theta(v-V^{\mathrm{simp.}}_{\mathrm{s.c.}}(\tilp,k_1,...,k_n)) \ .
\end{equation}
Within NNLL accuracy, and considering only independent emissions, this quantity is only non zero when there is a branch of the C/A clustering sequence, that is examined by the groomer as a result of all branches at wider angles being groomed away, and which contains two emissions labelled by $k_\alpha,k_\beta \in \{k_i\}$ which:
\begin{itemize}
\item would individually be groomed away
\item collectively pass the grooming condition
\item set a value of the observable larger than the cut on it: $V^{\mathrm{mMDT}}(\tilp,k_\alpha,k_\beta)>v$,
\end{itemize}
resulting in this pair of emissions being allowed by our simplified groomer when in fact they should be vetoed. Therefore, we evidently have $V^{\mathrm{mMDT}}(\tilp,k_1,..,k_\alpha,k_\beta,..,k_n)>v$ in this region of phase space and, as such, the clustering correction arises when the combination of step functions in Eq.~\eqref{eq:ClusteringThetas} is equal to $-1$, i.e. $V^{\mathrm{simp.}}_{\mathrm{s.c.}}(\tilp,k_1,..,k_\alpha,k_\beta,..,k_n)<v$. This region of phase space is illustrated on a Lund diagram in Fig.~\ref{fig:abelianClusteringPlane}.

We will now compute the clustering correction described above. To that end, we define the function:
\begin{multline}
\Theta_{\text{clust.}}^{\text{ind.}}\equiv \Theta(\theta_\alpha-\theta_{\alpha,\beta})\Theta(\theta_\beta-\theta_{\alpha,\beta}) \times \\ \Theta(\zc-z_\alpha)\Theta(\zc-z_\beta)\Theta(z_\alpha+z_\beta-\zc)\Theta(V_{\mathrm{s.c.}}(\tilp,k_\alpha,k_\beta)-v) \ ,
\end{multline}
where $\theta_\alpha$ ($\theta_\beta$) is the angle between the final-state quark and emission $\alpha$ ($\beta$), and $\theta_{\alpha,\beta}$ is the angle between partons $\alpha$ and $\beta$. Provided we ensure that the groomer examines the branch containing $\alpha$ and $\beta$, i.e. it is not stopped by a wider angle emission, $\Theta_{\text{clust.}}^{\text{ind.}}$ isolates the region of phase space which generates the clustering correction. The branch containing the two emissions responsible for the clustering correction is constrained by $\Theta(V_{\mathrm{s.c.}}(\tilp,k_\alpha,k_\beta)-v)$, which forms part of $\Theta_{\text{clust.}}^{\text{ind.}}$, to be at angles larger than $\left(v/\zc^b\right)^{1/a+b}$. The requirement on all other emissions that $V^{\mathrm{simp.}}_{\mathrm{s.c.}}(\tilp,k_1,...,k_n)<v$ forces all emissions other than $\alpha$ and $\beta$ which carry $z_i>\zc$ to be at angles smaller than $\left(v/\zc^b\right)^{1/a+b}$, thus guaranteeing that the groomer will examine the $\alpha,\beta$ branch, see Fig.~\ref{fig:abelianClusteringPlane}.

We can then write Eq.~\eqref{eq:ClusteringCorDef} as
\begin{multline}\label{eq:SigmaClustInd}
	\Sigma^{\text{clust.}}_{\text{ind.}}(v;\zc)=-\mathcal{H}(Q) \frac{1}{2!} \int [\sd  k_\alpha] [\sd  k_\beta] \mathcal{M}_{\mathrm{s.c}}^2(k_\alpha) \mathcal{M}_{\mathrm{s.c}}^2(k_\beta)  \Theta_{\text{clust.}}^{\text{ind.}} \\ \sum_{n=0}^\infty \frac{1}{n!}  \prod_{i=0}^{n}\int[\sd  k_i]\mathcal{M}_{\mathrm{c}}^2(k_i) \Theta\left(v-V^{\mathrm{simp}}_{\mathrm{s.c.}}(\tilp,k_1,...,k_n)\right) ,
\end{multline}
where $\mathcal{M}_{\mathrm{c}}^2$ was defined in Eq.~\eqref{eq:collinearME} with $\mathcal{M}_{\mathrm{s.c}}^2$ denoting its soft limit, the explicit form of which is given in eq. \eqref{eq:FclustInd}, and where $k_\alpha$ and $k_\beta$ now do not appear in the list of particles labelled $1$ to $n$ in the final line of Eq. \eqref{eq:SigmaClustInd}. The real emissions labelled with $i$, i.e. the ones not responsible for generating the clustering correction, can then be combined with $\mathcal{H}(Q)$ and evaluated using the techniques discussed in section \ref{sec:multiple_emissions} to give\footnote{In principle, $\mathcal{F}_{\text{clust.}}^{\text{ind.}}(v)$ is also a function of $\zc$ but turns out not to depend on $\zc$ within NNLL accuracy.} (for a single hemisphere)
\begin{equation}\label{eq:factorisedClust}
\Sigma^{\text{clust.}}_{\text{ind.},\ell}(v;\zc)= \mathcal{F}_{\text{clust.},\ell}^{\text{ind.}}(v) \exp[-R_{\mathrm{NLL},\ell}(v;\zc)] \ \ ,
\end{equation}
where 
\begin{multline}\label{eq:FclustInd}
	\mathcal{F}_{\text{clust.},\ell}^{\text{ind.}}(v)=- \frac{1}{2!}\int [\sd  k_\alpha][\sd k_\beta] \mathcal{M}_{\mathrm{s.c}}^2(k_\alpha) \mathcal{M}_{\mathrm{s.c}}^2(k_\beta)  \Theta_{\text{clust.}}^{\text{ind.}}\\
	=-C_F^2\int \frac{\alpha_s\big(k_{t\alpha}^2\big)}{2 \pi}\frac{\alpha_s\big(k_{t\beta}^2\big)}{2 \pi}\frac{2\sd  z_\alpha}{z_\alpha}\frac{2\sd  z_\beta}{z_\beta}\frac{\sd\phi}{2\pi }  \frac{\sd \theta_{\beta}^2}{\theta_\beta^2}\frac{\sd  \theta_{\alpha}^2}{\theta_{\alpha}^2} \Theta_{\text{clust}}^{\mathrm{ind.}} \Theta(\theta_\alpha-\theta_\beta)\ ,
\end{multline}
and we have used the symmetry between partons $\alpha$ and $\beta$ to eliminate $\frac{1}{2!}$ in favour of $\Theta(\theta_\alpha-\theta_\beta)$. For the jet mass, Eq.~\eqref{eq:factorisedClust} is just the NLL Sudakov factor, multiplied by the NLO clustering correction calculated in Ref.~\cite{Anderle:2020mxj}, where the argument of the coupling has been set to the transverse momentum of the emission.

Factorising the clustering correction from the Sudakov factor in this way allows one to write the full resummed integrated distribution as
\begin{equation}\label{eq:NLLResult}
\Sigma_{\mathrm{NNLL}}(v;\zc)=e^{-\bar{R}_{\mathrm{NNLL}}(v;\zc)}+(2 C_\ell^{\mathrm{r.c}}(v)+ 2\mathcal{F}_{\mathrm{clust.},\ell}(v))e^{-R_{\mathrm{NLL}}(v;\zc)} \ ,
\end{equation}
where we have anticipated a similar factorisation for the correlated emission clustering correction and grouped the clustering terms together as $\mathcal{F}_{\mathrm{clust.}}=\mathcal{F}_{\mathrm{clust.}}^{\mathrm{ind.}}+\mathcal{F}_{\mathrm{clust.}}^{\mathrm{cor.}}$. In Eq.~\eqref{eq:NLLResult} we have specified the NLL Sudakov factor where it multiplies $C^{\mathrm{r.c}}(v)$ and $\mathcal{F}_{\mathrm{clust.}}(v)$ so as to remove N$^3$LL terms which would be generated by the interplay of these term with the NNLL terms in the Sudakov factor.

Returning to evaluate the clustering correction, we can make two further approximations. Firstly, making the replacement $\alpha_s(k_{t\beta}^2)\to\alpha_s(k_{t\alpha}^2)$ is equivalent to integrating over the phase space of emission $\beta$ and dropping terms which are beyond NNLL accuracy.\footnote{We have checked this by expanding $\alpha_s(k_{t\beta}^2)\simeq\alpha_s\left(1+\beta_0\alpha_s\ln(\frac{k_{t\beta}^2}{Q^2})\right)$, evaluating the integrals over $z_\alpha,z_\beta, \phi$ and $\theta_\beta^2$ and observing that the logarithm of $k_{t\beta}$ eventually becomes a logarithm of $k_{t\alpha}$. Any additional terms not accounted for in our prescription are just constants with an additional power of $\alpha_s$ and so are N$^3$LL.} Secondly, we can drop the contribution of emission $\beta$ to the observable ($V(\tilp,k_\alpha,k_\beta)\to V(\tilp,k_\alpha)$) as the larger angle emission dominates the value of the observable. The latter results only in the neglect of terms which do not contain a logarithm of $v$, and so are N$^3$LL or are suppressed by powers of $\zc$, which are neglected.\footnote{We have explicitly checked this for the heavy hemisphere mass, and verified that as $\zc \to 0$ the result asymptotes to Eq.~\eqref{eq:abelianClust}}
We can now evaluate the integrals over $\theta_{\beta}$ and $\phi$ before exchanging the integral over $\theta^2_\alpha$ for one over $k_{t\alpha}^2$ to obtain
\begin{multline}
	\mathcal{F}_{\text{clust.},\ell}^{\text{ind.}}(v)=-\frac{4}{\pi}\text{Cl}_2\left(\frac{\pi}{3}\right) \bigg[ \int_{(v\zc^b)^{\frac{2}{a+b}}Q^2}^{\zc^2Q^2}\frac{\sd  k_{t\alpha}^2}{k_{t\alpha}^2} \left(\frac{\alpha_s\big(k_{t\alpha}^2\big)}{2\pi}\right)^2 \int_{k_{t\alpha}/Q}^{\zc}\frac{\sd  z_\alpha}{z_\alpha} \int_{\zc-z_\alpha}^{\zc}\frac{\sd  z_\beta}{z_\beta}   \\
	+\Theta(b) \int_{v^{\frac{2}{a}}Q^2}^{(v\zc^b)^{\frac{2}{a+b}}Q^2}\frac{\sd  k_{t\alpha}^2}{k_{t\alpha}^2} \left(\frac{\alpha_s\big(k_{t\alpha}^2\big)}{2\pi}\right)^2 \int_{k_{t\alpha}/Q}^{(k_{t\alpha}/Q)^{\frac{a+b}{b}} v^{\frac{-1}{b}}}\frac{\sd  z_\alpha}{z_\alpha}\int_{\zc-z_\alpha}^{\zc}\frac{\sd  z_\beta}{z_\beta}  \bigg]\ ,
\end{multline}
where we have made use of the soft and collinear parametrisation of the observable in terms of $a$ and $b$ given in eq. \eqref{eq:softColObservable} and the second line contributes only when $b>0$. The integral in the last line corresponds to the phase space indicated by the hashed region in figure \ref{fig:abelianClusteringPlane} and does not contribute at NNLL. It is therefore neglected from here on, and the remaining integrals evaluated, neglecting power corrections, to give
\begin{equation}\label{eq:abelianClust}
\mathcal{F}^{\text{ind.}}_{\text{clust.},\ell}(v)=-\left(\frac{\alpha_s C_F}{2\pi}\right)^2\frac{4\pi}{3}\text{Cl}_2\left(\frac{\pi}{3}\right)\frac{\ln 1/v}{a + b - 2 \lambda_v} +\text{N}^3\text{LL}\ ,
\end{equation}
where we have dropped power corrections and terms which are only enhanced by logarithms of $\zc$ and as such are N$^3$LL. 

As well as the NNLL $\mathcal{O}(\alpha_s^2)$ term, Eq.~\eqref{eq:abelianClust} also contains NNLL terms to all orders due to the running coupling which generates the term in the denominator proportional to $\alpha_s \beta_0\ln (v^{\frac{2}{a+b}})$. The running coupling also generates terms enhanced by logarithms of $\zc$ to all orders which are beyond NNLL accuracy and so are neglected.

We have checked eq. \eqref{eq:abelianClust} at $\mathcal{O}(\alpha_s^2)$ against \texttt{Event2} \cite{Catani:1996jh,Catani:1996vz} for the jet mass, width and Les Houches angularity and found good agreement. This check is possible because \texttt{Event2} provides event weights separated by colour factor and associates each momenta in the event with a species of parton. This allows one to construct both $V^{\mathrm{mMDT}}(\tilp,k_1,k_2)$ and $V^{\mathrm{simp}}(\tilp,k_1,k_2)$ for each event, and thus extract the clustering correction at $\mathcal{O}(\alpha_s^2)$.

At $\mathcal{O}(\alpha_s^2)$, eq. \eqref{eq:abelianClust} also agrees with the fixed order clustering correction for the jet mass given by eq. (35) in Ref. \cite{Anderle:2020mxj}. We also find agreement to all orders at NNLL accuracy with the corresponding terms in the SCET resummation of the groomed jet mass carried out in \cite{Frye:2016aiz}.

We have shown that a sub-jet consisting of two independent emissions, $\alpha$ and $\beta$, generates a correction with a single collinear logarithm of the observable, and that for this to happen
\begin{equation}
 z_\alpha,z_\beta<\zc \ \ \text{and} \ \ z_\alpha+z_\beta>\zc
\end{equation}
  must be satisfied. The generalisation of this to three emissions, $\alpha,\beta,\gamma$, is that 
\begin{equation}
\begin{split}
z_i<&\zc \\
z_i+z_j<&\zc\ ,\ \  i \neq j \\
z_\alpha+z_\beta+z_\gamma>&\zc \ ,
\end{split}
\end{equation}
where $i,j \in \alpha,\beta,\gamma$. As all three emissions are constrained to be at similar angles, the above configurations can have only one logarithm of the observable and so are N$^3$LL. A similar argument holds for more emissions so that at N$^n$LL accuracy one only has to consider clustering corrections involving up to $n$ emissions, the exception being NLL accuracy where there is no clustering correction. Therefore, the clustering correction result we give for the $C_F^2$ colour channel in Eq.~\eqref{eq:abelianClust} is sufficient for NNLL accuracy.

\subsection{Correlated emission clustering correction}

\begin{figure}
\centering
\includegraphics[width=0.45\textwidth]{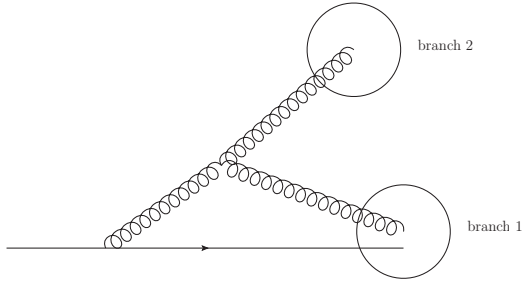}
\caption{The configuration responsible for the correlated emission clustering correction. Both gluons have $z_i<\zc$, but have a combined energy fraction greater than $\zc$. The two gluons are clustered in separate C/A branches, with branch two being de-clustered from the jet first and groomed away. The other gluon is then de-clustered from the quark and is also groomed away.}
\label{fig:CFCAClusterDiagram}
\end{figure}

The correlated emission clustering correction shares some similarities with the independent emission correction. The main difference is that it is due to a pair of emissions being groomed away in a region of phase space where the simplified groomer treats them as being retained, as opposed to the other way around for independent emissions. An $\mathcal{O}(\alpha_s^2)$ configuration where this occurs is shown in Fig.~\ref{fig:CFCAClusterDiagram}, which shows a pair of gluons, which are treated as being retained by the simplified groomer, but are actually removed due to the C/A clustering sequence. This configuration of emissions corresponds to the clustering correction computed in Ref. \cite{Anderle:2020mxj} for the $C_FC_A$ channel.

\begin{figure}
\centering
\includegraphics[width=0.45\textwidth,trim=360 550 100 60, clip]{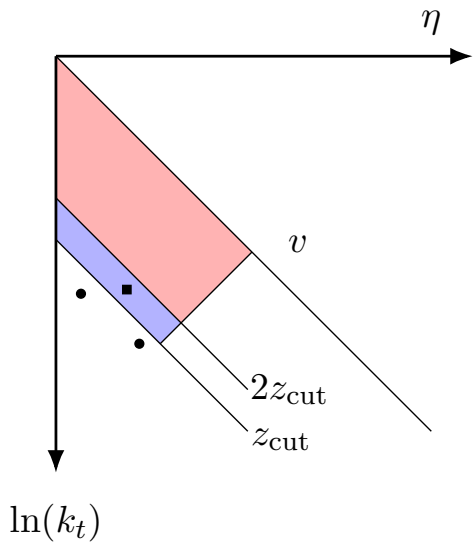}
\caption{Lund diagram showing the region of phase space responsible for the correlated clustering correction in blue, and in red the vetoed phase space responsible for the Sudakov factor. The black dots depict two correlated emissions, which, due to their separation in angle are clustered into different C/A branches, giving rise to the clustering correction. The black square represents the parent of these two emissions.}
\label{fig:NonAbelianClusteringPlane}
\end{figure}

We can calculate the correlated emission clustering correction up to NNLL accuracy by considering a pair of correlated emissions which we again label $\alpha$ and $\beta$. Configurations involving three emissions would be N$^3$LL as argued for the independent emission clustering correction. As per Eq.~\eqref{eq:ClusteringCorDef}, for correlated emissions the clustering correction is non vanishing if the correlated pair:
\begin{itemize}
	\item would individually be groomed away
	\item collectively pass the grooming condition
	\item set a value of the observable larger than the cut on it: $V^{\mathrm{mMDT}}(\tilp,k_\alpha,k_\beta)>v$  \ \ .
\end{itemize} 
The region of phase space responsible for the clustering correction is summarised in eq. \eqref{eq:encond_ind} with $V_{\mathrm{s.c.}}(\tilp,k_\alpha,k_\beta)>v$, which is illustrated in figure \ref{fig:NonAbelianClusteringPlane}. The relevant region of phase space for these two emissions is isolated by
\begin{multline}\label{eq:CorClustReg}
\Theta_{\text{clust.}}^{\text{cor.}} \equiv (\Theta(\theta_{\alpha,\beta}-\theta_\alpha)\Theta(\theta_{\beta}-\theta_\alpha)+(\alpha\leftrightarrow \beta))\times \\ \Theta(\zc-z_\alpha)\Theta(\zc-z_\beta)\Theta(z_\alpha+z_\beta-\zc) \Theta(V_{\mathrm{s.c.}}(\tilp,k_\alpha,k_\beta)-v)\ .
\end{multline}
In the same spirit as Eq.~\eqref{eq:SigmaClustInd}, we can then write:
\begin{multline}\label{eq:CorClustFact}
\Sigma^{\text{clust.}}_{\text{cor.},\ell}(v;\zc)= \mathcal{H}(Q) \frac{1}{2!} \int [\sd k_\alpha] [\sd k_\beta]\, \alpha^2_s(k_{t}^2) \, \bar{\mathcal{M}}_{\mathrm{cor}}^2(k_\alpha,k_\beta)  \Theta_{\text{clust.}}^{\text{cor.}} \\
\sum_{n=0}^\infty \frac{1}{n!}  \prod_{i=0}^{n}\int[\sd  k_i]\mathcal{M}_{\mathrm{c}}^2(k_i) \Theta(v-V^{\mathrm{simp.}}_{\mathrm{s.c.}}(\tilp,\{k_{i}\})),
\end{multline}
where $\bar{\mathcal{M}}_{\mathrm{cor}}^2(k_\alpha,k_\beta)$ is the squared matrix element for two correlated soft emissions, given in appendix A of \cite{Banfi:2018mcq}, but with the two factors of the coupling stripped off. The argument of both factors of the coupling is approximated as $k_t^2$, the total transverse momentum of the two partons \footnote{In the region of phase space relevant for the clustering correction, the relative transverse momenta of $a$ and $b$ must be of order the transverse momentum of the parent.} similar to what is done for the correlated emission correction in \cite{Banfi:2014sua,Banfi:2018mcq}. Like the independent emission case, it is not important exactly what the argument of both powers of the coupling are, only that they scale with the observable as $v^{\frac{2}{a+b}}Q^2$.

In the region of phase space given by Eq.~\eqref{eq:CorClustReg}, it is sufficient to use the collinear limit of $\bar{\mathcal{M}}_{\mathrm{cor}}^2(k_\alpha,k_\beta)$. Following the same steps as for the independent emission clustering correction, we can write now write eq. \eqref{eq:CorClustFact} as
\begin{equation}
\Sigma^{\text{clust.}}_{\text{cor.},\ell}(v;\zc)= \mathcal{F}_{\text{clust.},\ell}^{\text{cor.}}(v) \exp[-R_{\mathrm{NLL,\ell}}(v;\zc)] \ \ ,
\end{equation}
where,
\begin{equation}\label{eq:FclustCorDef}
\mathcal{F}_{\text{clust.},\ell}^{\text{cor.}}(v)=\frac{1}{2!}\int [\sd k_\alpha] [\sd k_\beta]\, \alpha^2_s(k_{t}^2) \, \bar{\mathcal{M}}_{\mathrm{cor}}^2(k_\alpha,k_\beta)  \Theta_{\text{clust.}}^{\text{cor.}}\ .
\end{equation}

We have not been able to compute these integrals analytically. However, it is possible to re-write them as an integral over the total transverse momentum of emissions $\alpha$ and $\beta$, which collects all the observable dependence and is straightforward to evaluate, multiplied by an integral which is computable numerically. We can write the phase space measure from Eq.~\eqref{eq:FclustCorDef} as
\begin{equation}\label{eq:NAPhaseSpace}
	[\sd k_\alpha] [\sd k_\beta] \sd k_t^2\delta_{k_t^2}
\end{equation}
where
\begin{equation}
	\delta_{k_t^2}=\delta\left(k_t^2 - (z_\alpha+z_\beta)^2\left(\frac{z_\alpha \theta_\alpha^2}{z_\alpha+z_\beta} + \frac{z_\beta \theta_\beta^2}{z_\alpha+z_\beta} - \frac{z_\alpha z_\beta \theta_{\alpha,\beta}^2}{(z_\alpha+z_\beta)^2}\right)\right)
\end{equation}
and $k_t$ is the total transverse momentum of $\alpha$ and $\beta$ in the collinear limit. As we are working with rIRC safe observables and up to NNLL accuracy we can make the replacement $V_{\mathrm{s.c.}}(\tilp,k_\alpha,k_\beta)\to V_{\mathrm{s.c.}}(\tilp,k_\alpha+k_\beta)$ \cite{Banfi:2004yd,Banfi:2014sua}. Moreover, in calculating $V_{\mathrm{s.c.}}(\tilp,k_\alpha+k_\beta)$, we can take the invariant mass of the $\alpha, \beta$ pair to be zero within NNLL accuracy. In the region of phase space defined by $\Theta_{\text{clust.}}^{\text{cor.}}$, the ordering $v<\zc$ then allows us to further simplify the limits on $k_t^2$ to $v^{\frac{2}{a+b}}\zc^{\frac{2b}{a+b}}Q^2<k_t^2<\zc^2Q^2$, where we have applied the parametrisation of the observable given in Eq.~\eqref{eq:softColObservable}.

We then choose $\theta_\alpha>\theta_\beta$ and provide a factor of two to account for the opposite ordering to write 
\begin{multline}\label{eq:FclustExpanded}
\mathcal{F}_{\text{clust.}}^{\text{cor.}}(v)=\\\int_{Q^2v^{\frac{2}{a+b}}\zc^{\frac{2b}{a+b}}}^{Q^2\zc^2} \frac{\sd  k_{t}^2}{k_{t}^2}\ \alpha_s^2(k_t^2) \times  \bigg(\int  [\sd k_\alpha] [\sd k_\beta] \mathcal{\bar{M}}_{\mathrm{cor}}^2(k_\alpha,k_\beta)\, k_t^2\, \delta_{k_t^2}\, \Theta(\theta_\alpha-\theta_\beta)\\\Theta(\theta_{\alpha,\beta}-\theta_\beta)\Theta(\zc-z_\alpha)\Theta(\zc-z_\beta)  \Theta(z_\alpha+z_\beta-\zc) \bigg) \ .
\end{multline}
After integrating over the delta function in Eq.~\eqref{eq:FclustExpanded}, the remaining integrals within $[\sd  k_\alpha] [\sd  k_\beta]$ can be re-scaled so that, other than the overall factor of $\frac{\sd  k_t^2}{k_t^2}$ and the argument of the coupling, there is no $k_t$ dependence in the integrals in Eq.~\eqref{eq:FclustExpanded}.  Leaving aside the integral over $k_t^2$, we can carry out the rest of the integrals numerically in the limit of $\zc\to 0$, to give \cite{Anderle:2020mxj}:
\begin{equation}\label{eq:NAclustering}
\mathcal{F}^{\text{cor.}}_{\text{clust.},\ell}(v)=\frac12 (C_FT_Rn_f1.754+C_FC_A1.161) \int_{Q^2v^{\frac{2}{a+b}}\zc^{\frac{2b}{a+b}}}^{Q^2\zc^2} \left(\frac{\alpha_s(k_t^2)}{2\pi}\right)^2 \frac{\sd  k_{t}^2}{k_{t}^2}  \ .
\end{equation}
This is then simply evaluated to give: 
\begin{equation}\label{eq:Corelated_Clust}
\mathcal{F}^{\text{cor.}}_{\text{clust.},\ell}= \left(\frac{\alpha_s }{2\pi}\right)^2 (C_F T_R n_f 1.754+C_FC_A1.161) \frac{\ln1/v}{a+b-2\lambda_v} \ ,
\end{equation}
where as for the independent emission clustering correction, we have omitted terms which are beyond our prescribed accuracy.
We re-iterate that the it is only important that the argument of the coupling scales with the observable so as to produce the factor of $\ln\left( v^{\frac{-2}{a+b}}\right)$ in the denominator of Eq.~\eqref{eq:Corelated_Clust}, and that the precise details of the argument do not matter at NNLL accuracy.

We have checked the leading $\mathcal{O}(\alpha_s^2)$ NNLL term in Eq.~\eqref{eq:Corelated_Clust} for the $C_FC_A$ channel using \texttt{Event2} for the jet mass, width, and Les Houches angularity (LHA), and found good agreement. The correlated clustering correction, and our resummed result as a whole, agrees with the collinear $\mathcal{O}(\alpha_s^2)$ calculation carried out for the jet mass in Ref. \cite{Anderle:2020mxj} and is consistent with the SCET resummation of the groomed jet mass \cite{Frye:2016aiz}, up to NNLL accuracy.

\section{Phenomenology}\label{sec:pheno}

So far, we have focused on the regime $v\ll\zc\ll1$ which mean that powers of $\zc$ have been ignored in the resummation. Nevertheless, this is not the only regime relevant for phenomenology. Firstly, one typically sets $\zc\sim 0.1$ in experimental analyses.\footnote{In tagging applications the optimal value of $\zc$ arises from considering the impact of the groomer on signal jets, for instance jets initiated by a boosted electroweak boson, as well as background QCD jets. For pure QCD studies, including testing resummed calculations against data, we are however free to investigate larger values of $\zc$ and our ability to describe them.} Therefore, terms of order $\alpha_s^{n} \zc \ln^n(v)$ could reasonably be expected to have a similar size to the NNLL terms which we resum. It is therefore desirable to resum these terms, as was done for the jet mass in Ref.~\cite{Marzani:2017mva}. We therefore show in Sect.~\ref{sec:FiniteZ} how our results can be modified so that such terms are resummed at the NLL level. By contrast, previous resummations of the groomed jet mass have either carried out the resummation at NLL accuracy, retaining power corrections in $\zc$ \cite{Marzani:2017mva}, or at NNLL accuracy but neglecting power corrections \cite{Frye:2016aiz}.

Although an interesting problem, the resummation of finite $\zc$ terms at the NNLL level, i.e. $\mathcal{O}(\alpha_s^n\zc\ln^{n-1}(v))$, is beyond the scope of this work. For typical values of $\zc\sim0.1$, such terms are likely to be small compared to the NNLL terms which survive in the small $\zc$ limit, but could be comparable in size to N$^3$LL corrections.
Secondly it is desirable to present results that are valid over the full experimentally measurable range of values of $v$, necessitating matching to fixed-order calculations to capture the limit where $v$  approaches $\zc$. We define a fixed-order matching prescription in Sect.~\ref{sec:matching} before studying, in Sect.~\ref{sec:Results}, the impact of including these terms for the jet mass, width and LHA.

\subsection{Finite \texorpdfstring{$\zc$}{zc}}\label{sec:FiniteZ}
We will initially deal with NLL resummation while retaining power corrections in $\zc$, as this will inform our matching procedure. The NLL resummation including these terms was carried out, for the jet mass distribution, in Ref.~\cite{Marzani:2017mva}. Dropping the arguments of the radiators for compactness, their result can be expressed as 
\begin{multline}
\frac{1}{\sigma_0}\frac{\sd \sigma}{\sd \ln\left(1/\rho\right)}=\begin{pmatrix} R^\prime_{q, \mathrm{NLL}} & R^\prime_{g, \mathrm{NLL}} \end{pmatrix}\\\exp \begin{pmatrix}
    -R_{q,\mathrm{NLL}}-R_{q\ \zc}-R_{q\to g} & R_{g\to q}\\
    R_{q\to g} & -R_{g,\mathrm{NLL}}-R_{g\ \zc}-R_{g\to q}\end{pmatrix} \begin{pmatrix}
    \sigma_q\\
    \sigma_g\end{pmatrix}\ ,
\end{multline}
where $\sigma_q$ and $\sigma_g$ are the Born cross-sections for quark and gluon initiated jets respectively, and $R_{g,\mathrm{NLL}}$ is the small $\zc$ radiator for gluon jets, which is related to the NLL quark radiator (see Eq. \eqref{eq:RPNLL}) by a change of colour factor from $C_F$ to $C_A$ and replacing the quark hard collinear anomalous dimension with the relevant one for gluons ($\gamma^0_{\mathrm{h.c,q}}\to\gamma^0_{\mathrm{h.c,g}}= -b_0$). Compared to Ref.~\cite{Marzani:2017mva} we have expressed the quark and gluon radiators as the small $\zc$ limit of the radiator $R^{\mathrm{NLL}}_{q/g}$, the NLL accurate radiator given by $g_1$ and $h_1$ in Appendix~\ref{app:NNLLSud}, plus the functions, $R_{q/g\ \zc}$ which vanish with $\zc$. This allows us to easily replace the parts of the radiators which survive in the small $\zc$ limit with the NNLL radiators given in Sect.~\ref{sec:simplified}. For a general observable $v$, the other components of the radiators are defined as follows: 
\begin{align}
	\nonumber
		R_{q \ \zc}(v,\zc) &= - \frac{C_F}{\pi}\bigg(\int_{1-\zc}^1 \sd z \, p_{gq}(z) +
		\int_0^{\zc} \sd z\, \left(p_{gq}(z)- \frac{2}{z}\right)  \bigg) \int_0^{z^2Q^2} \frac{\sd  k_t^2}{k_t^2} \alpha_s(k_t^2) \Theta\left(V_{\mathrm{sc}}(z,k_t)-v\right) \\ \nonumber
		R_{g \ \zc}(v,\zc) &= - \frac{C_A}{\pi} \int_{1-\zc}^1 \frac{2\, \sd z}{z} \int_0^{z^2 Q^2}\frac{\sd  k_t^2}{k_t^2}  \alpha_s(k_t^2) \, \Theta(V_{\mathrm{sc}}(z,k_t)-v) \\ \nonumber
		&- \int_0^{\zc} 2 \sd z \bigg(C_A \bigg(p_{gg}(z)-\frac{1}{z}-\frac{1}{1-z}\bigg)+T_Rn_f p_{qg}(z)\bigg) \int_0^{z^2 Q^2}\frac{\sd  k_t^2}{k_t^2} \frac{\alpha_s(k_t^2)}{\pi} \Theta\left(V_{\mathrm{sc}}(z,k_t)-v\right)\\ \nonumber
		R_{q\to g}(v,\zc) &= \frac{C_F}{\pi}\int_{1-\zc}^1 \sd z\,  p_{gq}(z) \int_0^{z^2 Q^2} \frac{\sd  k_t^2}{k_t^2} \alpha_s(k_t^2) \, \Theta(V_{\mathrm{sc}}(z,k_t)-v) \\
		R_{g\to q}(v,\zc) &= \frac{T_R n_f }{\pi} \int_{1-\zc}^1 2 \, \sd z\,  p_{qg}(z)\int_0^{z^2 Q^2}  \frac{\sd  k_t^2}{k_t^2} \alpha_s(k_t^2)\,  \Theta(V_{\mathrm{sc}}(z,k_t)-v) \  \ .
\end{align}

To complete the matching of our NNLL result to the NLL finite $\zc$ result we must also include the NNLL corrections which are not exponentiated. As we do not attempt to capture finite $\zc$ NNLL terms it is sufficient to make sure that however these are included, we reproduce Eq.~\eqref{eq:NLLResult} on taking the small $\zc$ limit.
For our purposes $\sigma_g=0$, and we can normalize our distributions to the Born cross-section for the production of a $q\bar{q}$ pair and finally write the NNLL result including NLL finite $\zc$ effects as: 
\begin{multline}
\Sigma_{\mathrm{NNLL}}(v;\zc)=\begin{pmatrix} 1+2 C_\ell^{\mathrm{r.c}}(v)+ 2 \mathcal{F}_{\mathrm{clust.},\ell}(v) &, 1 \end{pmatrix} \\ \times \exp \begin{pmatrix}
    -\bar{R}_{q,\mathrm{NNLL}}-R_{q\ \zc}-R_{q\to g} & R_{g\to q}\\
    R_{q\to g} & -R_{g,\mathrm{NLL}}-R_{g\ \zc}-R_{g\to q}\end{pmatrix}\begin{pmatrix}
    1\\
    0\end{pmatrix}\ ,
\end{multline}
where $\mathcal{F}_{\mathrm{clust.}}=\mathcal{F}^{\mathrm{clust.}}_{\mathrm{ind.}}+\mathcal{F}^{\mathrm{clust.}}_{\mathrm{cor.}}$ and $\bar{R}_{q,\mathrm{NNLL}}$ is given in Appendix~\ref{app:NNLLSud}. In the small $\zc$ approximation the functions $R_{i\ \zc}$ and $R_{i\to j}$ vanish, where $i\neq j$ can be $q$ or $g$ and we return to our NNLL result. 

\subsection{Matching}\label{sec:matching}
We now turn to matching our resummed calculation to NLO fixed-order calculations. Although our NNLL calculation captures all logarithms of the observable at $\mathcal{O}\big(\alpha_s^2\big)$, which are not suppressed by powers of $\zc$, there still exist NLO terms which are $\mathcal{O}(\zc)$ but diverge as $v\to 0$. For this reason we use a multiplicative matching procedure similar to that employed in Ref. \cite{Caletti:2021oor}. This choice of matching scheme ensures that the $\mathcal{O}(\zc)$ logarithmically divergent terms are suppressed by the Sudakov factor as $v\to 0$, whilst correctly capturing both the NNLL resummed distribution at small $\zc$, the finite $\zc$ terms at NLL accuracy and the full fixed-order distribution at NLO.

We introduce the notation $\Sigma_{\mathrm{NNLL}}^{(1)}(v)$ for the $\mathcal{O}(\alpha_s)$ part of the resummed distribution, with $\Sigma_{\mathrm{NNLL}}^{(2)}(v)$ representing the $\alpha_s^2$ terms. The matched cumulative distribution then reads
\begin{multline}\label{eq:matching}
\Sigma(v)=\Sigma_{\mathrm{NNLL}}\left[1+ \left(\Sigma^{(1)}(v)-\Sigma_{\mathrm{NNLL}}^{(1)}(v)\right) + \right.\\\left. \left(\Sigma^{(2)}(v)-\Sigma_{\mathrm{NNLL}}^{(2)}(v)\right) - \Sigma_{\mathrm{NNLL}}^{(1)}(v)\left(\Sigma^{(1)}(v)-\Sigma_{\mathrm{NNLL}}^{(1)}(v)\right) \right].
\end{multline}
The first term in Eq.~\eqref{eq:matching} is just the NNLL distribution. The term, $\big(\Sigma^{(1)}(v)-\Sigma_{\mathrm{NNLL}}^{(1)}(v)\big)$, is the $\mathcal{O}(\alpha_s)$ part of the distribution with the terms already captured by the all orders NNLL distribution subtracted. This term is multiplied by the full NNLL distribution thus generating spurious terms at $\mathcal{O}\big(\alpha_s^2\big)$, which are removed by the last term, $\Sigma_{\mathrm{NNLL}}^{(1)}(v)\left(\Sigma^{(1)}(v)-\Sigma_{\mathrm{NNLL}}^{(1)}(v)\right)$. The term, $\big(\Sigma^{(2)}(v)-\Sigma_{\mathrm{NNLL}}^{(2)}(v)\big)$ is the $\mathcal{O}\big(\alpha_s^2\big)$ part of the distribution with the terms already captured by the NNLL distribution subtracted. There are no spurious terms within our accuracy generated by the interplay of this term with $\Sigma_{\mathrm{NNLL}}$.

We perform the matching for the process $e^+e^-\to q \bar{q}$ for three jet shapes: the heavy hemisphere mass, the width $\lambda_1^{\mathrm{WTA}}$, and the Les Houches angularity $\lambda_{0.5}^{\mathrm{WTA}}$. The fixed-order results are calculated, for $\zc=0.1$, using \texttt{Event2} with the jet clustering, and grooming being done using fast-jet \cite{Cacciari:2011ma}. Each event was partitioned into two hemispheres by a plane perpendicular to the C/A jet axis,\footnote{This partitioning of the hemispheres, compared to using the thrust axis does not have an impact on the NNLL results for groomed observables, only on the fixed order matching.} mMDT is then run on each hemisphere and the jet shape calculated on the groomed hemispheres. The larger value of the observable from the two hemispheres is then binned.

\subsection{Results}\label{sec:Results}

\begin{figure}
\centering
\begin{subfigure}{0.475\textwidth}
\centering
\includegraphics[width=1\textwidth]{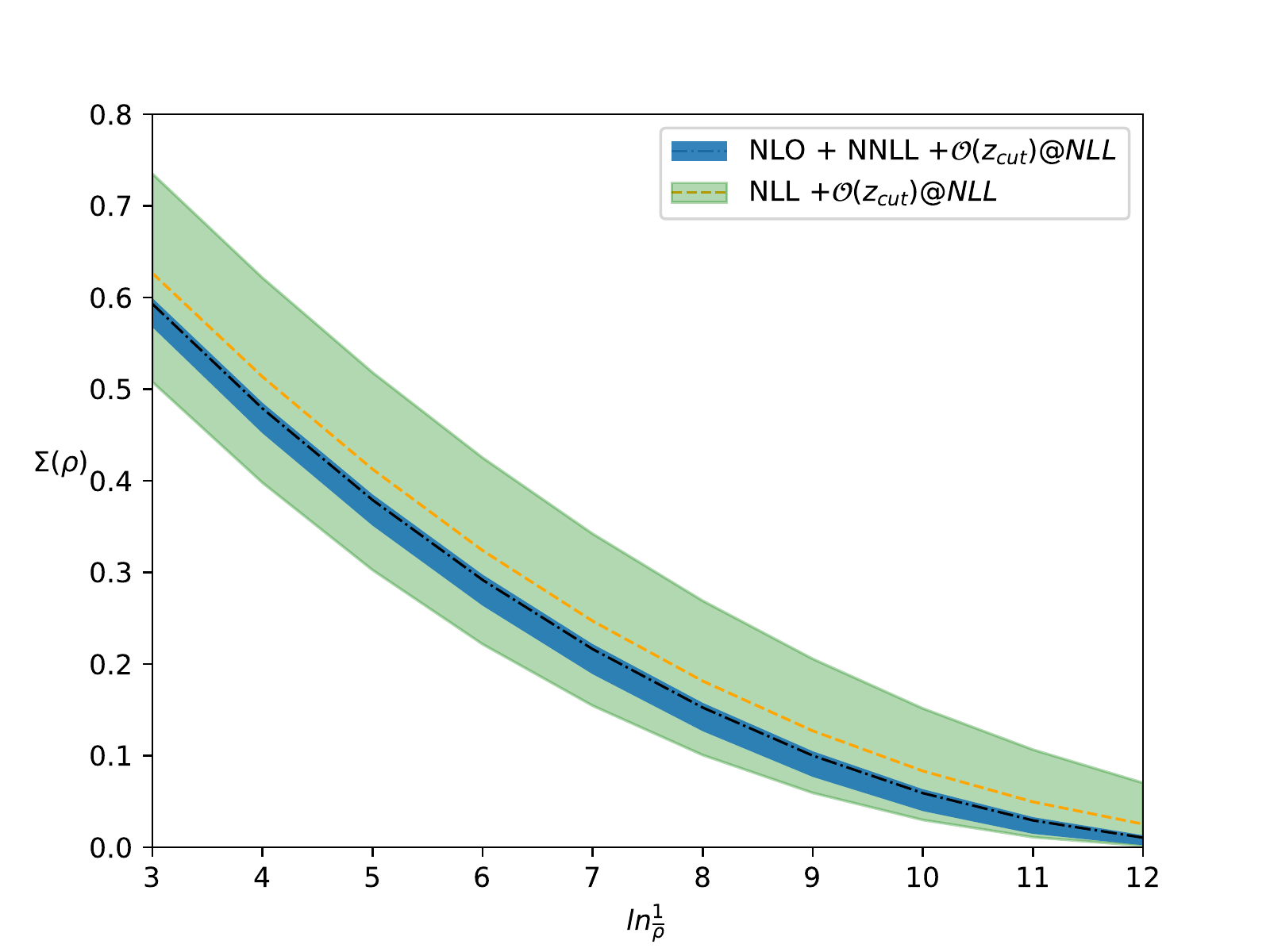}
\caption{$\rho$ (Jet Mass)}
\label{fig:JM_Uncert_Zc01}
\end{subfigure}
\begin{subfigure}{0.475\textwidth}
\centering
\includegraphics[width=1\textwidth]{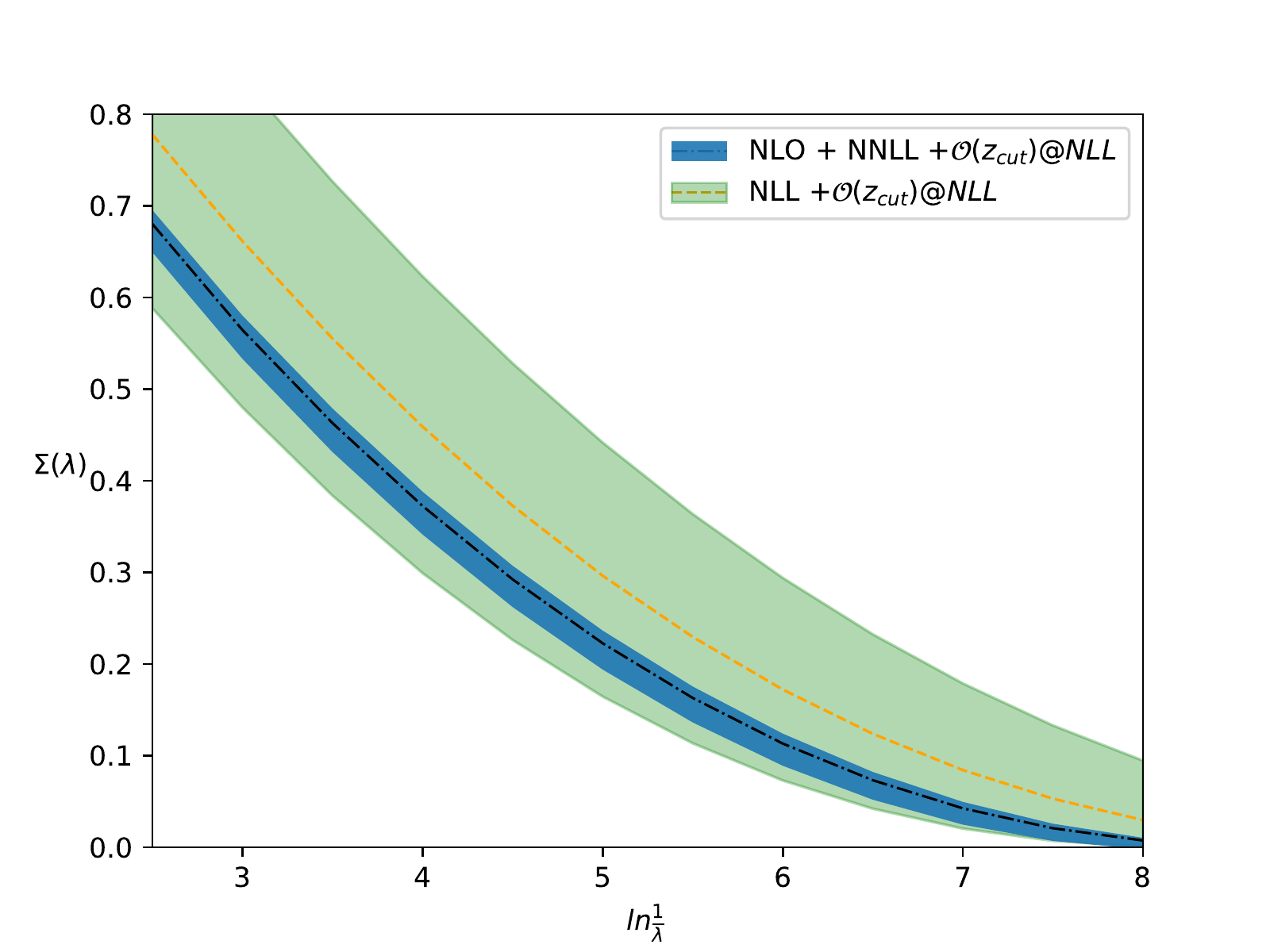}
\caption{$\lambda^{\mathrm{WTA}}_1$}
\label{fig:A1_Uncert_zc01}
\end{subfigure}
\begin{subfigure}{0.475\textwidth}
\centering
\includegraphics[width=1\textwidth]{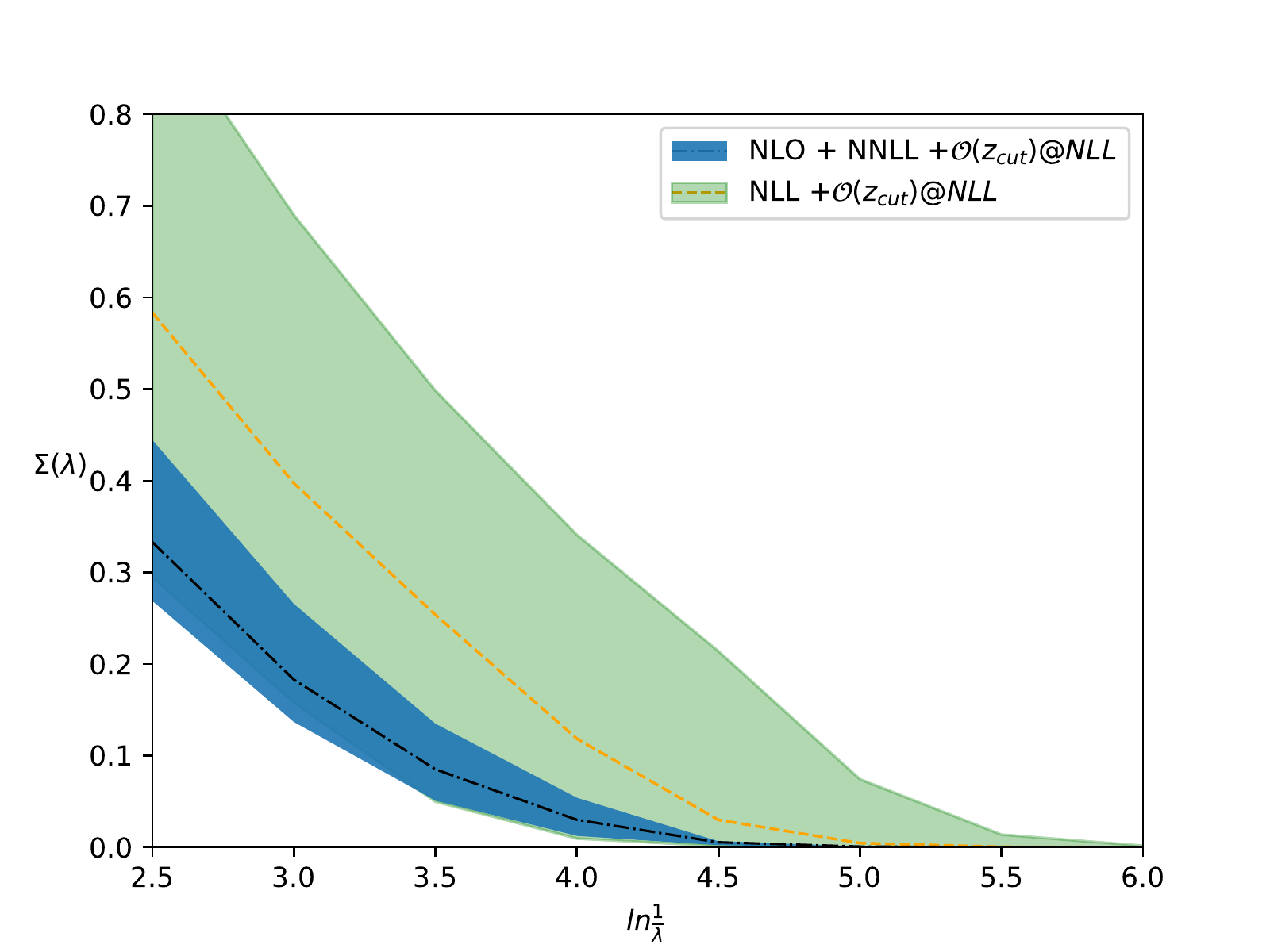}
\caption{$\lambda^{\mathrm{WTA}}_{0.5}$}
\label{fig:A05_Uncert_zc01}
\end{subfigure}
\caption{Matched NNLL predictions alongside the NLL result, both with finite $\zc$ effects showing the central values and uncertainty bands.}
\label{fig:NLLvsLLUncert}
\end{figure}

Having defined the matching procedures for incorporating both the $\mathcal{O}(\zc)$ effects at NLL and the full NLO distribution, we now investigate the impact of including these effects and the size of the uncertainties, which we derive from resummation and renormalisation scale variation. Figure \ref{fig:NLLvsLLUncert} shows our matched results for the heavy hemisphere mass, width, and Les Houches angularity, alongside the respective NLL with finite $\zc$ results for a typical value of $\zc=0.1$ so as to show the size of the NNLL corrections. 

For the jet mass and width, the uncertainty bands are found by simultaneously varying the resummation and renormalisation scales by factors of two in such a way as to avoid introducing any spurious terms within the stated accuracy, similar to Ref.~\cite{Napoletano:2018ohv}. The renormalisation scale enters through the argument of the coupling, and is included by making the replacement 
\begin{equation}
\alpha_s(Q)\to\alpha_s(xQ)+\alpha_s^2(Q)\beta_0 \ln(x) \ \ ,
\end{equation} 
and varying $x$ between $0.5$ and $2$. This preserves the NNLL result whilst generating our renormalisation scale variation. The resummation scale uncertainty is found for the small $\zc$ NNLL calculation by making the replacement $R(v,\zc) \to R_{\mathrm{NLL}}(x v, \zc)-R_{\mathrm{NLL}}'(v,\zc)\ln x+R^{\mathrm{remainder}}(xv, \zc)$, where again, $x$ is varied between $0.5$ and $2$. Here $R^{\mathrm{NLL}}(v,\zc)$ contains purely the next to leading logarithms in the small $\zc$ limit, and nothing else, and $R^{\mathrm{remainder}}(v, \zc)=R(v,\zc)-R_{\mathrm{NLL}}(v,\zc)$. This has the effect of introducing terms proportional to $\ln2$ which are N$^3$LL and higher without introducing spurious terms within the accuracy that we control, thus giving an estimate of the possible size of missing N$^3$LL terms. To obtain the resummation scale uncertainty for the NLL accurate results\footnote{This includes the finite $\zc$ corrections which we match our NNLL result to.} shown in figure \ref{fig:NLLvsLLUncert} we simply replace $v\to xv$ which introduces terms due to scale variation at NNLL. For the NNLL result matched to the NLL finite $\zc$ resummation, the prescriptions for NNLL and NLL scale variation detailed above are applied simultaneously to the relevant terms so that the scale variation starts at NNLL for terms suppressed by powers of $\zc$, and N$^3$LL for terms which survive in the small $\zc$ limit.

The Les Houches angularity ($\lambda^{\mathrm{WTA}}_{0.5}$) becomes sensitive to very small transverse momentum emissions much faster than the other two observables. It was therefore necessary to introduce a freezing scale for the coupling, which we set at $1$ GeV to avoid divergences due to the Landau pole. The uncertainty band shown in Fig.~\ref{fig:A05_Uncert_zc01} for this observable therefore incorporates variation in the freezing scale by factors of two as well as the uncertainties previously discussed.

\begin{figure}
\centering
\begin{subfigure}{0.475\textwidth}
\centering
\includegraphics[width=1\textwidth]{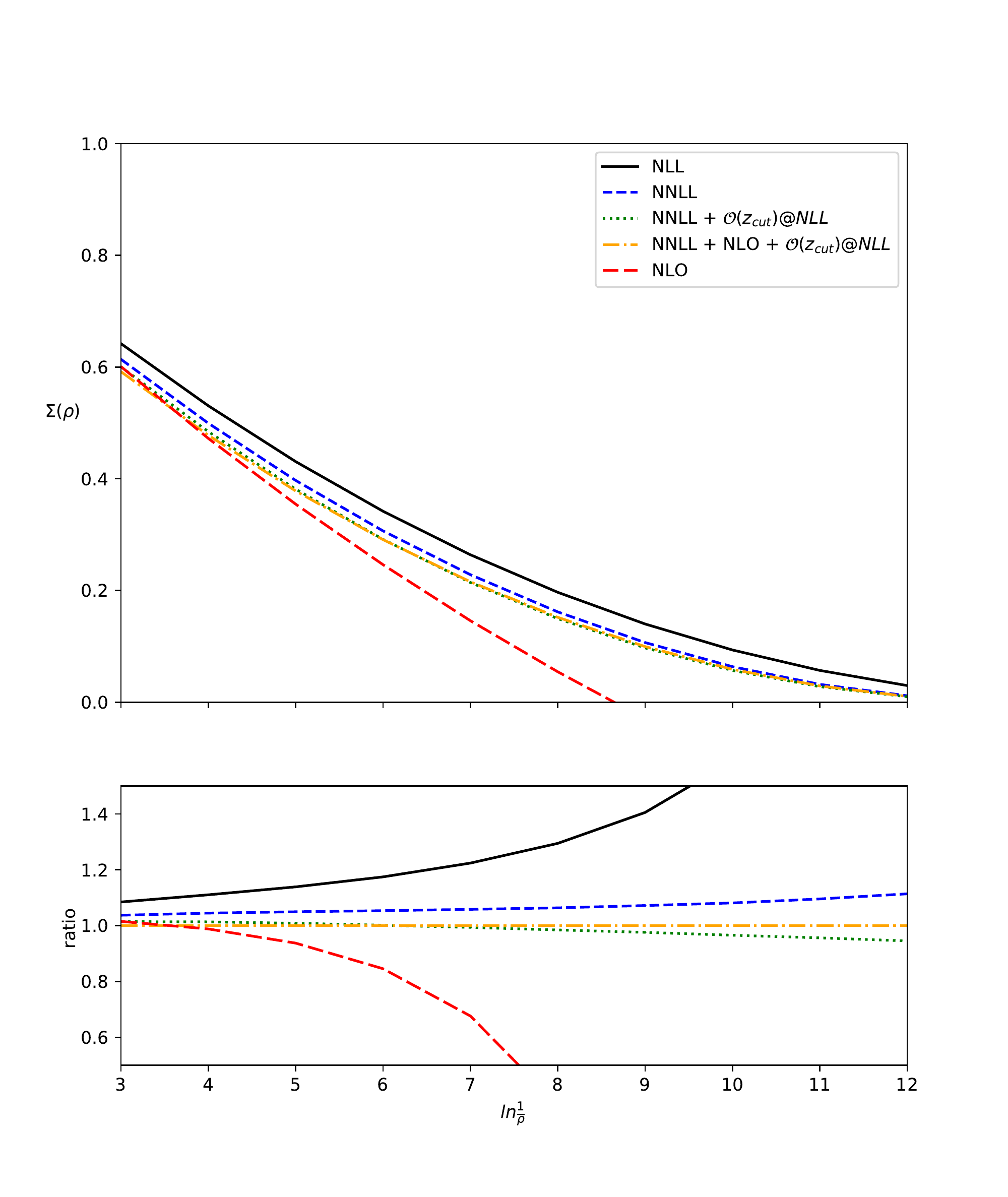}
\caption{$\zc=0.1$}
\label{fig:JMzc01Comparison}
\end{subfigure}
\begin{subfigure}{0.475\textwidth}
\centering
\includegraphics[width=1\textwidth]{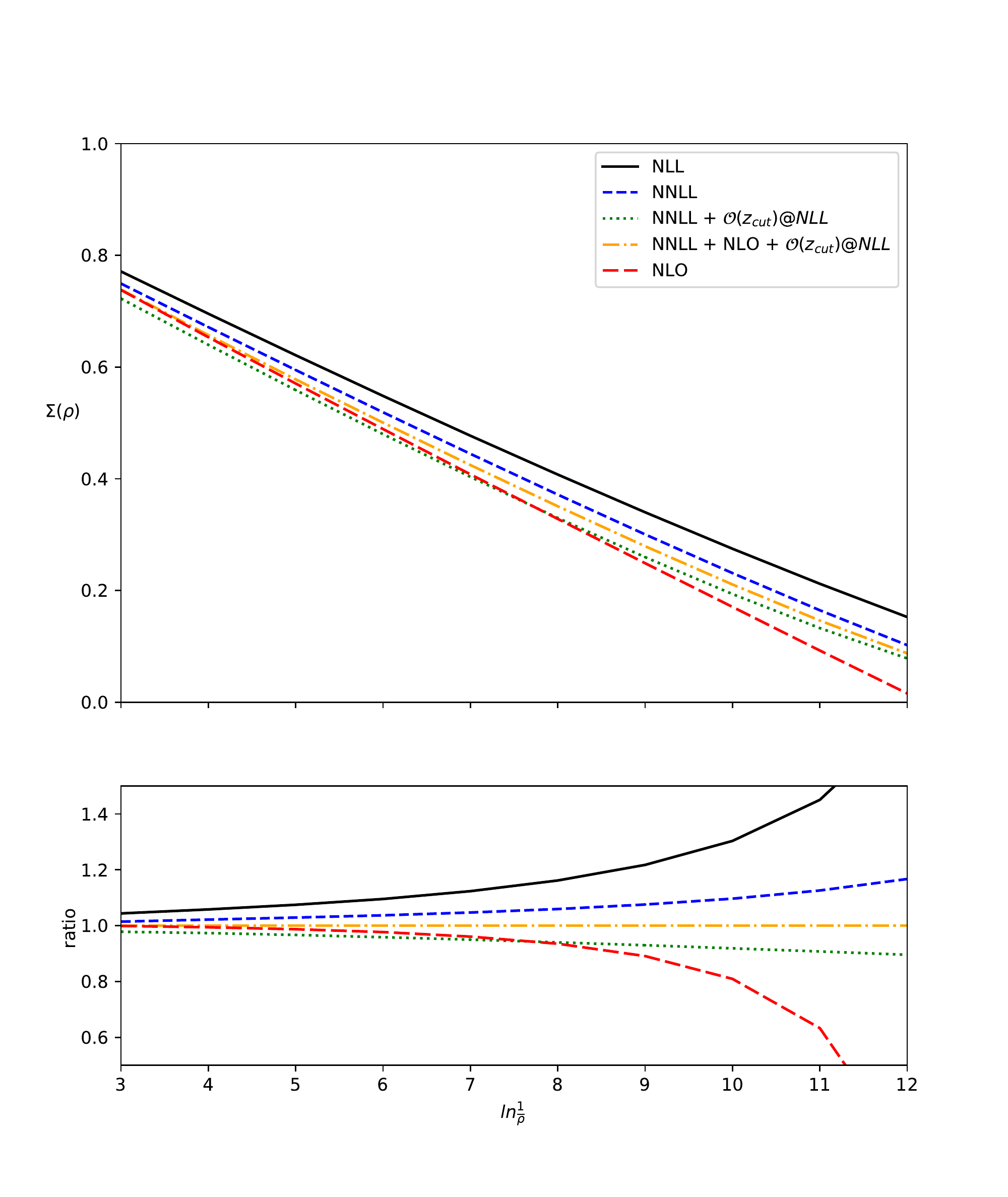}
\caption{$\zc=0.2$}
\label{fig:JMzc02Comparison}
\end{subfigure}
\begin{subfigure}{0.475\textwidth}
\centering
\includegraphics[width=1\textwidth]{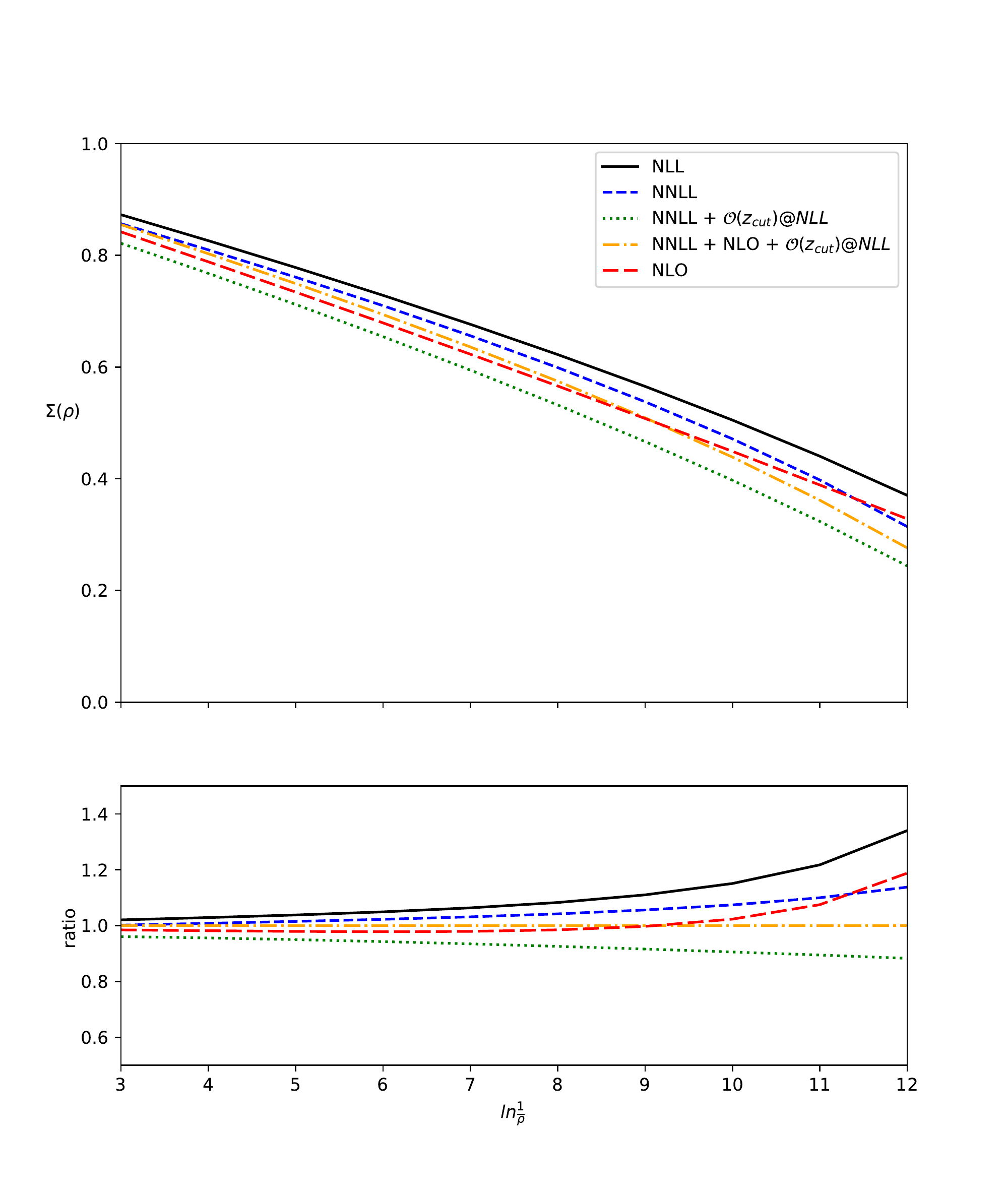}
\caption{$\zc=0.3$}
\label{fig:JMzc03Comparison}
\end{subfigure}
\caption{Predictions for the cumulative jet mass distribution at various levels of accuracy for three different values of $\zc$}
\label{fig:JM_ZCcomparison}
\end{figure}

From Fig.~\ref{fig:NLLvsLLUncert} we see that the inclusion of NNLL terms is important both for the noticeable shift in the central value, but also for the large reduction in the uncertainty it provides. The importance of this reduction is reinforced by Ref. \cite{Reichelt:2021svh} which commented on the relatively large theoretical uncertainties in the NLL results which were used in their phenomenological study. The reduction in uncertainty obtained by going to NNLL accuracy brings the theoretical uncertainty close to the size of experimental uncertainties shown in Ref. \cite{Reichelt:2021svh}.  

Figure \ref{fig:JM_ZCcomparison} shows results for the jet mass for $\zc=0.1,0.2,0.3$ at several levels of accuracy alongside the ratio of these results to our NNLL result matched to NLO and the $\mathcal{O}(\zc)$ NLL calculation. Particularly as $\zc$ is increased, we can see that at moderate values of $\ln(v)$ resumming finite $\zc$ terms at NLL becomes as important as NNLL resummation of small $\zc$ terms. For values of $\zc\simeq 0.1$ or larger, which might be used for phenomenology, finite $\zc$ effects at NLL accuracy should therefore be retained, if one is keeping similarly sized NNLL terms in the limit $\zc \to 0$. One caveat to this is that, if the impact of resumming finite $\zc$ effects is investigated on results which are matched to fixed order, the size of the finite $\zc$ effects may well appear reduced, as the finite $\zc$ effects will automatically be included up to the order in $\alpha_s$ which the matching is carried out to. This may explain the small effect of finite $\zc$ resummation observed in previous studies carried out at NLL accuracy \cite{Marzani:2017mva,Larkoski:2020wgx}.

\section{Conclusions}\label{sec:conclusion}

In this work the distribution of additive rIRC safe observables, computed on jets groomed with mMDT, in the context of $e^+e^-$ collisions, was calculated at NNLL accuracy  in the small $\zc$ limit. Motivated by the calculations presented in Ref. \cite{Anderle:2020mxj}, the resummation is structured as an inclusive piece, defined in section \ref{sec:NNLL_Structure}, which is added to a clustering correction, which accounts for the effect of the C/A clustering sequence. Our result agrees up to NNLL accuracy with a previous resummation of the groomed jet mass carried out in SCET \cite{Frye:2016aiz,Kardos:2020ppl}, which provides a powerful cross check between the two results.  However, there are differences between the two calculations starting at N$^3$LL. 

Having presented our NNLL result in the small $\zc$ limit, we then modified our result so as to include NLL terms which vanish with $\zc$, which were previously resummed in \cite{Dasgupta:2013ihk,Marzani:2017mva}. This was done because NLL terms suppressed by a power of $\zc$ could reasonably be expected to be numerically similar in size to the NNLL terms we resummed in the small $\zc$ limit, for values of $\zc\sim0.1$. We then performed fixed order matching to $\mathcal{O}(\alpha_s^2)$ for the heavy jet mass, width, and Les Houches angularity. Our results for these three observables are shown in section \ref{sec:Results} at different levels of accuracy to assess the impact of different effects. From these results we saw that, for values of $\zc\geq0.1$, the NLL finite $\zc$ effects are indeed of a similar size to the NNLL terms kept in the small $\zc$ limit, and so, for phenomenology, should be retained if one is keeping the NNLL terms. Although we noted that if matching to NLO fixed order calculations, the effect of resumming finite $\zc$ may well be less noticeable.

One continuation of this work would be to extend the resummation presented here to gluon jets. Together with the work presented here, this could then be used to produce NNLL accurate predictions for a range of observables which could be compared with measurements carried out at the LHC. NLL finite $\zc$ effects and NLO matching could be included in such predictions in the same way as we have done in this work. As discussed in Sect.~\ref{sec:Results}, going to NNLL accuracy brings the uncertainties close to the size of experimental uncertainties shown in Ref.~\cite{Reichelt:2021svh}. With these reduced uncertainties it may then be feasible to perform an extraction of $\alpha_s$ from jet substructure measurements as investigated in Ref. \cite{Hannesdottir:2022rsl}, but using a range of groomed observables. This could help with breaking the degeneracy between fitting the fraction of quark and gluon jets simultaneously with $\alpha_s$ that was noted in \cite{Proceedings:2018jsb}.

Though this work has only considered additive observables, we believe that it should be possible to extend the methods presented here to non-additive observables such as the broadening \cite{Rakow:1981qn}. Another possibility is extending the resummation to soft drop with $\beta>0$. These extensions would further expand the range of observables for which NNLL accurate predictions could be produced for hadron collider phenomenology.

\acknowledgments
We would like to thank Aditya Pathak for help in understanding aspects of the SCET resummation of the groomed jet mass, for the purposes of comparison of our results with the existing results for jet mass in the literature.  We also thank Gregory Soyez for discussions related to the subject of this paper. This work has been funded by the European Research Council (ERC) under the European Union's Horizon 2020 research and innovation program (grant agreement n.o. 788223) (MD, JH and BKE) and by the U.K.'s Science and Technologies Facilities Council via grant ST/T001038 (MD) and via a PhD studentship award which funded JH during the course of this work.

\appendix

\section{Exponentiation of soft emissions.}\label{app:softExponentiation}

Here we provide an explicit demonstration that all emissions satisfying $z_i<\zc$ can be dropped from $V^{\mathrm{simp}}_{\mathrm{s.c.}}(\tilp,k_1,...,k_n)$  up to power corrections in $v$, regardless of whether or not they are groomed away and thus can be fully exponentiated.\\
We start by considering an ensemble of soft and collinear emissions which are ordered in angle with an emission, $(s)$, singled out as the one which stops the groomer, in configurations in which the groomer does not remove all emissions:
\begin{multline}\label{eq:start}
\Sigma^{\text{real}}=\sum_{p=0}^\infty\sum_{m=0}^p \bigg\{ \prod_{j=0}^m \bar{\alpha}\int_0^{\zc}\frac{\sd  z_j}{z_j} \int_{\theta_s^2}^1 \frac{\sd  \theta_j^2}{\theta_j^2} \Theta(\theta_j-\theta_{j+1})\\
\bigg( \bar{\alpha}\int_0^{\zc}\frac{\sd  z_s}{z_s} \int_0^1 \frac{\sd  \theta_s^2}{\theta_s^2}\Theta(\zc-z_i) +\bar{\alpha}\int_{\zc}^{1}\frac{\sd  z_s}{z_s} \int_0^1 \frac{\sd  \theta_s^2}{\theta_s^2} \Theta(v-V_{\mathrm{s.c.}}(\tilp,k_s)-V_{\mathrm{s.c.}}(\tilp,\{k_i \})) \bigg) 
\\ \prod_{i=0}^{p-m} \bigg( \bar{\alpha}\int_0^{\zc}\frac{\sd  z_i}{z_i} \int_0^{\theta_s^2} \frac{\sd  \theta_i^2}{\theta_i^2} +  \bar{\alpha}\int_{\zc}^{1}\frac{\sd  z_i}{z_i} \int_0^{\theta_s^2} \frac{\sd  \theta_i^2}{\theta_i^2} \bigg)\Theta(\theta_i-\theta_{i+1}) \bigg\} \ ,
\end{multline}
where $\bar{\alpha}=\frac{C_F\alpha_s}{\pi}$. The term on the second line  has $z_s>\zc$ and represents all configurations where the groomer is stopped, and thus all emissions at smaller angles (those on the final line) are retained and contribute to the observable. This term can be re-written as
\begin{multline}\label{eq:SoftEmissionCorrection}
\sum_{p=0}^\infty\sum_{m=0}^p \bigg\{ \prod_{j=0}^m \bar{\alpha}\int_0^{\zc}\frac{\sd  z_j}{z_j} \int_{\theta_s^2}^1 \frac{\sd  \theta_j^2}{\theta_j^2} \Theta(\theta_j-\theta_{j+1})
\left(\bar{\alpha}\int_{\zc}^{1}\frac{\sd  z_s}{z_s} \int_0^1 \frac{\sd  \theta_s^2}{\theta_s^2} \Theta(v-V_{\mathrm{s.c.}}(\tilp,k_s)-V_{\mathrm{s.c.}}(\tilp,\{k_i \})) \right) \\ 
\prod_{i=0}^{p-m} \bigg( \bar{\alpha}\int_0^{\zc}\frac{\sd  z_i}{z_i} \int_0^{\theta_s^2} \frac{\sd  \theta_i^2}{\theta_i^2} +  \bar{\alpha}\int_{\zc}^{1}\frac{\sd  z_i}{z_i} \int_0^{\theta_s^2} \frac{\sd  \theta_i^2}{\theta_i^2} \bigg)\Theta(\theta_i-\theta_{i+1}) \bigg\} =\\
\int\frac{\sd \nu}{2\pi i\nu} e^\nu \sum_{p=0}^\infty\sum_{m=0}^p \bigg\{ \prod_{j=0}^m \left(\bar{\alpha}\int_0^{\zc}\frac{\sd  z_j}{z_j} \int_{\theta_s^2}^1 \frac{\sd  \theta_j^2}{\theta_j^2} \Theta(\theta_j-\theta_{j+1})\right)\bar{\alpha}\int_{\zc}^{1}\frac{\sd  z_s}{z_s} \int_0^1 \frac{\sd  \theta_s^2}{\theta_s^2}e^{-\nu \frac{V_{\mathrm{s.c.}}(\tilp,k_s)}{v}} \\ \prod_{i=0}^{p-m} \bigg(\bar{\alpha}\int_{\zc}^{1}\frac{\sd  z_i}{z_i} \int_0^{\theta_s^2} \frac{\sd  \theta_i^2}{\theta_i^2}e^{-\nu \frac{V_{\mathrm{s.c.}}(\tilp,k_i)}{v}}+ \bar{\alpha}\int_0^{\zc}\frac{\sd  z_i}{z_i} \int_0^{\theta_s^2} \frac{\sd  \theta_i^2}{\theta_i^2}+\\
\bar{\alpha}\int_0^{\zc}\frac{\sd  z_i}{z_i} \int_0^{\theta_s^2} \frac{\sd  \theta_i^2}{\theta_i^2}(e^{-\nu \frac{V_{\mathrm{s.c.}}(\tilp,k_i)}{v}}-1)\bigg)\Theta(\theta_i-\theta_{i+1}) \bigg\}.
\end{multline}
where the step functions involving the observable have been written in their Laplace representation. Emissions at smaller angles (labelled with $i$) are now represented by three terms, one for emissions with $z>\zc$, which always contribute to the observable, another accounting for emissions with $z<\zc$ as if they do not contribute to the observable, and a correction (the final line of Eq. \eqref{eq:SoftEmissionCorrection}) accounting for the fact that emissions with $z<\zc$ do in fact contribute to the observable. Emissions at smaller angles than $\theta_s$ can be exponentiated and the aforementioned correction evaluated:
\begin{multline}\label{eq:ExponentiatedSoftEmissionCorrection}
=\int\frac{\sd \nu}{2\pi i\nu} e^\nu \sum_{q=0}^\infty \bigg\{ \prod_{j=0}^q \bigg(\bar{\alpha}\int_0^{\zc}\frac{\sd  z_j}{z_j} \int_{\theta_s^2}^1 \frac{\sd  \theta_j^2}{\theta_j^2} \Theta(\theta_j-\theta_{j+1})\bigg) \bar{\alpha}\int_{\zc}^{1}\frac{\sd  z_s}{z_s} \int_0^1 \frac{\sd  \theta_s^2}{\theta_s^2}e^{-\nu \frac{V_{\mathrm{s.c.}}(\tilp,k_s)}{v}} \\ \exp\bigg[\bar{\alpha}\int_{\zc}^{1}\frac{\sd  z}{z} \int_0^{\theta_s^2} \frac{\sd  \theta^2}{\theta^2}e^{-\nu \frac{V_{\mathrm{s.c.}}(\tilp,k)}{v}} + \bar{\alpha}\int_0^{\zc}\frac{\sd  z}{z} \int_0^{\theta_s^2} \frac{\sd  \theta^2}{\theta^2}+\mathcal{O}(\bar{\alpha_s}\zc^a\theta_s^{a+b})\bigg] \bigg\}.
\end{multline}
One can consider expanding this exponential to any order and observe that correction term, the $\mathcal{O}(\bar{\alpha_s}\zc^a\theta_s^{a+b})$ terms on the final line of the above equation, will at most generate N$^3$LL power corrections in both $v$ and $\zc$ once the integral over $\theta_s$ is carried out, which is sufficient to show that this term can be dropped. Equivalently, any emission softer than $\zc$ can be dropped from $V_{\mathrm{s.c.}}(\tilp,k_1,...,k_n)$, as we have made use of in the main text.

\section{The Sudakov radiator at NNLL}\label{app:NNLLSud}
In this appendix we evaluate the integral in Eq.~\eqref{eq:Sudakov1}. The only point to note here is the fact that, at NNLL accuracy, we need to retain the difference between the physical energy of the emission, $z$, and the light-cone variables $(z^{(1)},z^{(2)})$ of Eq.~\eqref{eq:Sudvar}. As is customary, we arrange the Sudakov radiator in terms of functions of definite logarithmic accuracy, as follows:
\begin{align}\label{eq:nnllsud}
	R_{\text{NNLL},\ell}(v;\zc) = - g_{1,\ell}(\lambda_v;\lambda_{\zc}) - g_{2,\ell}(\lambda_v,\lambda_{\zc}) - h_2 (\lambda_v) - h_3(\lambda_v) \ \ ,
\end{align}
where
\begin{align}
	g_{1,\ell}(\lambda_v;\lambda_{\zc}) &= \frac{C_F}{\pi \alpha_s \beta_0^2} \lambda_{\zc} \ln\left(1-\frac{2 \lambda_v}{a+b_\ell}\right) \label{eq:g1} \ \ , \\\nonumber
	g_{2,\ell}(\lambda_v;\lambda_{\zc}) &= \frac{C_F}{\pi \alpha_s \beta_0^2} \lambda_{\zc}^2 \left(\frac{a-2 \lambda_v}{a+b_\ell - 2 \lambda_v}\right)  + \frac{C_F \beta_1}{\pi \beta_0^3} \lambda_{\zc} \frac{2 \lambda_v + (a+b_\ell) \ln \left(1-\frac{2 \lambda_v}{a+b_\ell}\right)}{a+b_\ell - 2\lambda_v} \\
	&-\frac{C_F}{\pi^2 \beta_0^2} \lambda_{\zc} \frac{\lambda_v\, K_{\text{CMW}}} {a+b_\ell - 2 \lambda_v} - \frac{C_F \alpha_s}{2\pi} \frac{\pi^2}{6} \label{eq:g2} \ \ , \\
	h_{1,\ell}(\lambda_v) &=  \frac{C_F \gamma_{\mathrm{h.c}}^{(0)}}{2\pi \beta_0} \ln\left(1-\frac{2 \lambda_v}{a+b_\ell}\right) \label{eq:h2}\ \ , \\ \nonumber
	h_{2,\ell}(\lambda_v) &= \frac{C_F \alpha_s \gamma_{\mathrm{h.c}}^{(0)} \,\beta_1}{2\pi \beta_0^2 (a + b_\ell - 2\lambda_v)}\left( (a+b_\ell) \ln\left(1-\frac{2 \lambda_v}{a+b_\ell}\right) + 2\lambda_v \right) \\
	&-  \frac{C_F \alpha_s \gamma_{\mathrm{h.c}}^{(1)} }{2\pi^2 \beta_0 (a+b_\ell - 2\lambda_v)} \lambda_v \label{eq:h3}\ \ .
\end{align}
We finally note that consistent with our logarithmic accuracy we are free to re-expand the pure $\pi^2/6$ constant in $g_{2,\ell}$. We include this term as part of the definition of $C_\ell^{\mathrm{r.c}}$ in Eq.~\eqref{eq:C1}. We make such a choice to avoid un-wanted N$^3$LL contributions. Therefore, we define the following function which is employed in our master resummed formula:
\begin{align}\label{eq:nnllbarsud}
	\bar{R}_{\text{NNLL},\ell}(v;\zc) = - g_{1,\ell}(\lambda_v;\lambda_{\zc}) - \bar{g}_{2,\ell}(\lambda_v,\lambda_{\zc}) - h_2 (\lambda_v) - h_3(\lambda_v) \ \ ,
\end{align}
where
\begin{align}
	\nonumber
	\bar{g}_{2,\ell}(\lambda_v;\lambda_{\zc}) &= \frac{C_F}{\pi \alpha_s \beta_0^2} \lambda_{\zc}^2 \left(\frac{a-2 \lambda_v}{a+b_\ell - 2 \lambda_v}\right)  + \frac{C_F \beta_1}{\pi \beta_0^3} \lambda_{\zc} \frac{2 \lambda_v + (a+b_\ell) \ln \left(1-\frac{2 \lambda_v}{a+b_\ell}\right)}{a+b_\ell - 2\lambda_v} \ \ .
\end{align}
\newpage
\bibliographystyle{JHEP}
\bibliography{mMDT_Resummation_2022_v2.bib}

\end{document}